\documentclass[useAMS,usenatbib]{mnras}

\usepackage{hyperref}
\usepackage{graphicx}
\usepackage{multirow}
\usepackage{verbatim}

\title[GAMA/G10 CSED]{Galaxy And Mass Assembly: the evolution of the cosmic spectral energy distribution from $z = 1$ to $z = 0$ \thanks{Herschel is an ESA space observatory with science instruments provided by European-led Principal Investigator consortia and with important participation from NASA.}}
\author[Andrews et al.]{S. K. Andrews$^1$\thanks{E-mail: stephen.andrews@icrar.org}, S. P. Driver$^{1, 2}$, L. J. M. Davies$^1$, P. R. Kafle$^1$, A.~S.~G. Robotham$^1$, \newauthor K. Vinsen$^1$, A. H. Wright$^{1, 3}$, J. Bland-Hawthorn$^4$, N. Bourne$^5$, M. Bremer$^6$, \newauthor E. da Cunha$^7$, M. Drinkwater$^8$, B. Holwerda$^9$, A.~M. Hopkins$^{10}$, L.~S. Kelvin$^{11}$, \newauthor J. Loveday$^{12}$, S. Phillipps$^6$, S. Wilkins$^{12}$ \\
$^1$International Center for Radio Astronomy Research, The University of Western Australia, 35 Stirling Highway, Crawley, WA 6009, Australia \\
$^2$School of Physics \& Astronomy, University of St Andrews, North Haugh, St Andrews, KY16 9SS, UK \\
$^3$Argelander-Institut f\"{u}r Astronomie, Universit\"{a}t Bonn, Auf dem H\"{u}gel 71, 53121 Bonn, Germany \\
$^4$Sydney Institute for Astronomy, School of Physics A28, University of Sydney, NSW 2006, Australia \\
$^5$Institute for Astronomy, University of Edinburgh, Royal Observatory, Blackford Hill, Edinburgh EH9 3HJ, UK \\
$^6$H H Wills Physics Laboratory, University of Bristol, Tyndall Avenue, Bristol, BS8 1TL, UK \\
$^7$Research School of Astronomy and Astrophysics, Australian National University, Canberra, ACT 2611, Australia \\
$^8$School of Mathematics and Physics, University of Queensland, QLD 4072, Australia \\
$^9$Department of Physics and Astronomy, University of Louisville, 102 Natural Science Building, Louisville, KY 40292, USA \\
$^{10}$Australian Astronomical Observatory, PO Box 915 North Ryde, NSW 1670, Australia \\
$^{11}$Astrophysics Research Institute, Liverpool John Moores University, IC2, Liverpool Science Park, 146 Brownlow Hill, Liverpool, L3 5RF, UK \\
$^{12}$Astronomy Centre, Department of Physics and Astronomy, University of Sussex, Falmer, Brighton BN1 9QH, UK \\
}
\begin{document}

\nocite{*}
\date{Accepted 1988 December 15. Received 1988 December 14; in original form 1988 October 11}

\pagerange{\pageref{firstpage}--\pageref{lastpage}} \pubyear{2002}

\maketitle

\label{firstpage}

\begin{abstract}
We present the evolution of the Cosmic Spectral Energy Distribution (CSED) from $z = 1 - 0$. Our CSEDs originate from stacking individual spectral energy distribution fits based on panchromatic photometry from the Galaxy and Mass Assembly (GAMA) and COSMOS datasets in ten redshift intervals with completeness corrections applied. Below $z = 0.45$, we have credible SED fits from 100 nm to 1 mm. Due to the relatively low sensitivity of the far-infrared data, our far-infrared CSEDs contain a mix of predicted and measured fluxes above $z = 0.45$. Our results include appropriate errors to highlight the impact of these corrections. We show that the bolometric energy output of the Universe has declined by a factor of roughly four -- from $5.1 \pm 1.0$ at $z \sim 1$ to $1.3 \pm 0.3 \times 10^{35}~h_{70}$~W~Mpc$^{-3}$ at the current epoch. We show that this decrease is robust to cosmic variance, SED modelling and other various types of error. Our CSEDs are also consistent with an increase in the mean age of stellar populations. We also show that dust attenuation has decreased over the same period, with the photon escape fraction at 150~nm increasing from $16 \pm 3$ at $z \sim 1$ to $24 \pm 5$ per cent at the current epoch, equivalent to a decrease in $A_\mathrm{FUV}$ of 0.4~mag. Our CSEDs account for $68 \pm 12$ and $61 \pm 13$ per cent of the cosmic optical and infrared backgrounds respectively as defined from integrated galaxy counts and are consistent with previous estimates of the cosmic infrared background with redshift.
\newline
\newline
\newline
\end{abstract}

\begin{keywords}
galaxies: general; galaxies: evolution; cosmic background radiation; cosmology: observations; 
\end{keywords}

\section{Introduction}
\label{sec:intro}

The Cosmic Spectral Energy Distribution (CSED, e.g. \citealt{driver08,driver15,dominguez11,gilmore12,somerville12}), describes the total energy generated as a function of wavelength for a cosmologically representative volume at some specified time. This is different to the photon budget, which describes the photons passing through the same volume at that time, and the extragalactic background light (EBL, e.g. \citealt{dwek98,hauser01}), which describes the radiation received per unit solid angle. One can define the resolved CSED which arises from discrete sources, and the total CSED which includes both discrete and diffuse light. This work measures the resolved CSED, which we will refer to as simply ``the CSED" unless indicated.

The CSED encodes statistical information about the ongoing processes of galaxy formation and evolution. This link becomes obvious when one considers the (resolved) CSED to be the sum of the individual spectral energy distributions (SEDs) of all galaxies in some cosmologically representative volume. The optical to near-infrared portion of a galaxy's SED encodes information about stellar mass, both gas and stellar phase metallicity, and the ages of stellar populations (e.g. \citealt{taylor11,madau14}), while the ultraviolet and total infrared emission can be used to estimate a galaxy's current star formation rate (e.g. \citealt{kennicutt98,madau14,davies16}). About half the energy produced by stars is attenuated by dust within their galaxy and re-radiated in the mid- and far-infrared, with the shape and magnitude of a galaxy's far-infrared SED depending on dust mass, temperature, geometry and composition (e.g. \citealt{dacunha08,dunne11,symeonidis13}). Finally, emission from an active galactic nucleus can also shape a galaxy's SED, and this contribution becomes increasingly significant at higher redshifts (see e.g. \citealt{richards06}). Analysing the total CSED allows for the extraction of the population average, weighted by energy and number density, of key quantities within the specified volume. Tracing the evolution of the CSED with lookback time therefore allows a reconstruction of the evolution of the energy output from stars, dust and AGN.

The integrated galactic light (IGL), which represents the resolved component of the EBL, can be determined from the redshifted CSED using a volume-weighted integral over all time. To derive the EBL from the IGL, one should consider additional contributions from diffuse radiation from the epoch of reionization \citep{cooray12b} and faint intra-halo light \citep{cooray12,zemcov14}. Recent measurements \citep{hess13,magic16,driver16b} suggest the diffuse components of the EBL represent an excess of approximately 20 per cent over the IGL in the near-infrared. This excess is marginally  significant given uncertainties in the IGL (dominated by cosmic sample variance) and EBL measurements.

The EBL and IGL have historically received more interest than the full, panchromatic CSED (e.g. \citealt{partridge67a,partridge67b,franceschini08,finke10,dominguez11,inoue13,driver16b}). Multiple groups have also examined how the cosmic infrared background at specific wavelengths builds up as a function of redshift, which if combined is equivalent to deriving the far-infrared portion of the CSED (e.g. \citealt{marsden09,jauzac11,berta11,bethermin12b,viero13}). Measurements of the CSED -- an instantaneous quantity, as opposed to an integrated measurement over all of Universal history -- at multiple epochs have a greater ability to constrain cosmological models of galaxy formation. 

Multiple groups (e.g. \citealt{bethermin10,carniani15,fazio04,gardner00,madau00,xu05}) have measured the IGL, but these efforts are generally restricted to one region of the electromagnetic spectrum \citep{driver16b}. To study the CSED from the far-ultraviolet to the far-infrared requires a combination of sufficiently deep and wide multiwavelength imaging and spectroscopic datasets and a means of deriving consistent photometry across the wavelength range. With the aid of SED fitting tools, the CSED and IGL can be characterized over the wavelength range while simultaneously avoiding systematic errors induced by inhomogenous data reduction methods.

The Galaxy And Mass Assembly survey \citep{driver11,liske15} is ideally suited to measuring the recent time evolution of the CSED. GAMA is an integrated multiwavelength imaging and spectroscopic campaign to examine the distribution of matter and energy on kpc to Mpc scales in the low-redshift ($z ~< 0.25$) Universe. GAMA covers 180 square degrees of the equatorial sky in three fields to a spectroscopic completeness of $\sim$98 per cent. An intermediate redshift ($0.5 ~< z ~< 1$) analogue to GAMA (in sample size and wavelength coverage) was assembled by \citet{davies15} and \citet{andrews16} using existing public data in the Cosmological Origins Survey (COSMOS; \citealt{scoville07}) field. This work aims to characterise the evolution in the CSED using a combination of the GAMA and COSMOS datasets.

Three previous measurements of the CSED at low redshifts ($z < 0.2$) have been made using the GAMA dataset. \citet{driver12} calculated the CSED using the GAMA far-ultraviolet to $K_s$ luminosity functions, and extrapolated it to the far-infrared using the \citet{dale02} models of dust attenuation. \citet{kelvin14} measured the CSED as a function of galaxy morphology for $z < 0.06$. More recently, \citet{driver15} measured the CSED and its evolution over the redshift range $0 < z < 0.2$ by fitting SEDs to a cosmologically representative sample of galaxies. Here we improve on the latter CSED measurement in two respects: i) using an updated photometric catalogue \citep{wright16a} with improved deblending and reduced gross photometric errors, and ii) expanding the redshift range using re-reduced publicly available data in the COSMOS field \citep{andrews16}. These CSEDs complement the recent IGL measurement by \citet{driver16b}. In Section \ref{sec:data}, we describe the multiwavelength data sets, techniques used to derive photometry and the SED modelling techniques we use to interpolate between these photometric measurements. Then in Section \ref{sec:csed}, we determine the CSED and examine its reliability. In Section \ref{sec:conclusion}, we present concluding remarks.

We use AB magnitudes and assume $H_0 = 70~h_{70}~\mathrm{km~s^{-1}~Mpc^{-1}}$, $\Omega_\Lambda = 0.7$ and $\Omega_m = 0.3$ throughout.

\section{Data}
\label{sec:data}

\begin{figure*}
\begin{minipage}{7in}
\begin{center}
\includegraphics[width=0.99\linewidth]{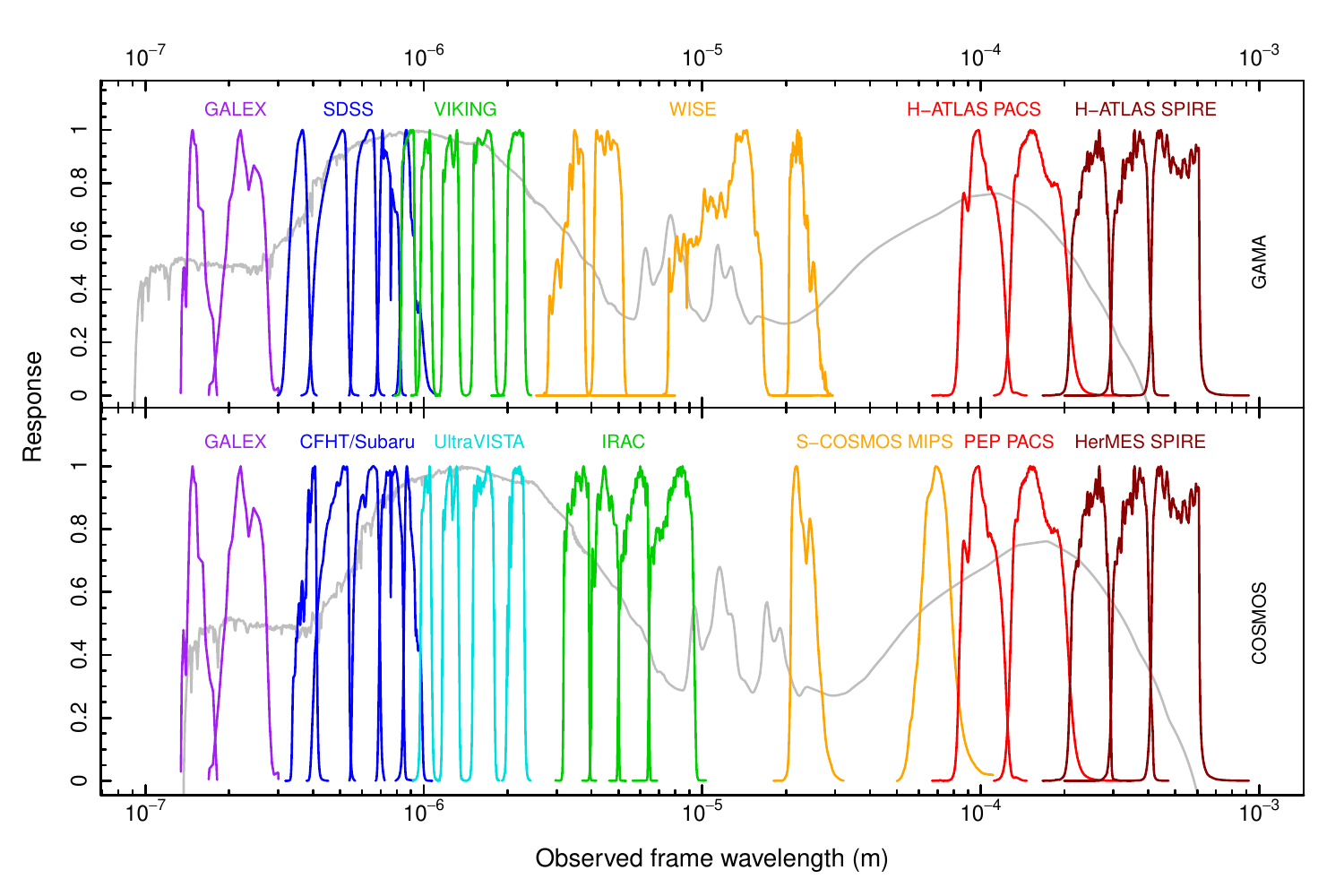}
\caption{The combined filter curves for the GAMA equatorial (top) and COSMOS (bottom) data sets. Also plotted is the \citet{driver12} model CSED (constructed from the cosmic star formation history), which illustrates what the energy-weighted average galaxy SED may look like at $z=0$ (top) and $z=0.5$ (bottom).}
\label{fig:filters}
\end{center}
\end{minipage}
\end{figure*}

\subsection{GAMA equatorial regions}
The GAMA \citep{driver11, liske15} spectroscopic campaign targeted 180 square degrees of the equatorial sky in three fields, centered on R.A.=9h, 12h and 14.5h, using the AAOmega spectrograph mounted on the 3.9~m Anglo-Australian Telescope. GAMA obtained redshifts for $\sim$200k galaxies to a depth of $r < 19.8$~mag (SDSS Petrosian). This is complemented by ultraviolet imagery from the \textit{Galaxy Evolution Explorer} (\textit{GALEX}; \citealt{martin05}), optical data from the Sloan Digital Sky Survey (SDSS DR7; \citealt{abazajian09}), near-infrared data from VISTA (Visible and Infrared Telescope for Astronomy) Kilo-degree Infrared Galaxy survey (VIKING; \citealt{edge13}), mid-infrared data from the \textit{Wide-field Infrared Survey Explorer} (\textit{WISE}; \citealt{wright10}) and far-infrared data from \textit{Herschel}-Atlas \citep{eales10}, as summarized by \citet{driver15}. The combined grasp of the GAMA filters is shown in Figure \ref{fig:filters} (upper).

\citet{wright16a} implemented a novel program, the Lambda-Adaptive MultiBand Deblending Algorithm in R (\textsc{lambdar}), which is capable of deriving (forced) aperture matched photometry from the far-ultraviolet to the far-infrared. \textsc{lambdar} is explicitly designed to correctly deblend objects in the far-infrared, where resolution and sensitivity are the lowest. Particular care was taken in building a set of robust aperture definitions in order to obtain an optimal photometric solution. Deblended fluxes are then obtained using \textsc{lambdar} across all 21 bands using these aperture definitions convolved with the point spread function. This approach minimises the potential for gross photometric inconsistencies, including table and aperture mismatches. The resulting catalogue, \textit{LDRSummaryPhotometryv01}, contains consistent flux and error measurements for approximately 220,000 sources across the GAMA wavelength range. \citet{wright16a} demonstrate that using this catalogue leads to clear improvements in SED fits and star formation rate estimators over table matching.

\subsection{G10/COSMOS}
The G10 region (R.A. = 149.55\degr--150.65\degr, Dec = +1.80--+2.73\degr) is a subset of the broader COSMOS region chosen for its relatively high spectroscopic completeness ($\sim45\%$ for extra-galactic sources with $i^+ < 22$~mag). It forms an intermediate-redshift comparison to GAMA. \citet{davies15} combined existing spectra from the PRIsm MUlti-object Survey (PRIMUS; \citealt{coil11,cool13}), the VIMOS-VLT Deep Survey (VVDS; \citealt{garilli08}) and SDSS DR10 \citep{ahn14}, re-reduced zCOSMOS \citep{lilly07,lilly09} spectra and photometric redshifts \citep{ilbert09}, obtaining reliable spectroscopic redshifts for over 22000 sources. This is complemented by publicly available ultraviolet images from \textit{GALEX} \citep{zamojski07}, optical images from the Canada-France-Hawaii and Subaru telescopes \citep{capak07,taniguchi07,taniguchi16} and near-infrared data from UltraVISTA \citep{mccracken12}. In the mid-infrared, data is available from the \textit{Spitzer} survey in COSMOS (S-COSMOS; \citealt{sanders07}) and the \textit{Spitzer} Large Area Survey with HyperSuprimeCam (SPLASH; Capak et al. in prep), while far-infrared data has been published by PACS (Photodetector Array Camera and Spectrometer) Evolution Probe (PEP; \citealt{lutz11}) and the \textit{Herschel} Multitier Extragalactic Survey (HerMES; \citealt{oliver12}). The combined filter curve for these datasets is shown in Figure \ref{fig:filters} (lower).

\citet{andrews16} used an equivalent procedure to \citet{wright16a} to derive consistent flux measurements in 38 bands and photometric redshifts from this dataset for approximately 186,000 sources brighter than $i^+ < 25$~mag. \citet{andrews16} demonstrated that using their catalogues results in goodness of SED fits being equivalent to or better than those derived from the recent COSMOS2015 catalogue \citep{laigle16}. 

\subsection{SED modelling}
\label{sec:seds}

\citet{driver17} fitted SEDs to the GAMA and G10/COSMOS data using the spectral analysis code \textsc{magphys} (Multi-wavelength Analysis of Galaxy Physical Properties; \citealt{dacunha08}). These models are computed using the \citet{bc03} stellar libraries, a \citet{chabrier03} initial mass function, and the \citet{charlot00} prescription for dust attenuation which consists of a two-component description of the interstellar medium (stellar birth clouds and diffuse interstellar medium). The dust emission is computed via energy balance -- the energy absorbed by dust in the ultraviolet to near-infrared range is re-emitted in the mid- to far-infrared range using empirically-calibrated dust emission components: polycyclic aromatic hydrocarbons, warm dust emitting in the mid-infrared, and two components of dust in thermal equilibrium with varying temperatures. \textsc{magphys} was modified to use the upper limits in the \citet{andrews16} and \citet{wright16a} catalogues, increase the range of possible dust masses and increase the photometric error floor from 5 to 10 per cent. 

\citet{driver17} demonstrate that the SED fits are generally robust and obtain stellar and dust mass densities and cosmic star formation rates from $z = 5$ to $z = 0$ in line with literature estimates. However, the fitting procedure and underlying imaging data may leave systematic impacts on measurements of the CSED:

\begin{itemize}
\item The \textsc{magphys} templates do not, as yet, incorporate AGN emission. This affects predominantly the higher redshift end of the sample.
\item Dust properties are poorly constrained or extrapolated for a significant portion of the combined sample due to the low depth and resolution of the far-infrared data. This is especially important for the high-redshift end of the sample, where many objects only have adoped upper limits in the far-infrared.
\item The GAMA sample does not have 70~$\mu$m imaging. This results in a near-complete inability to constrain the warm dust properties of individual galaxies.
\end{itemize}

We address these potential systematic effects in Section \ref{sec:csedbad}.

From the GAMA catalogue, we select all galaxies with good spectroscopic redshifts (NQ $> 2$ from \textit{SpecObjv27}) and coverage in FUV, NUV and 250~$\mu$m, which play an important role in constraining star formation and dust properties. This reduces the sample area to 129 deg$^2$ and $\sim$147k objects. AGN contamination in the GAMA sample is assumed to be negligible (see Section \ref{sec:csedbad}), with the relevant \citet{driver17} cut removing only 32 objects. From the G10/COSMOS catalogue, we select all galaxies in non-flagged regions (as denoted by SG\_MASTER = 0 and MASK\_COSMOS2015 = 0 from \textit{G10CosmosLAMBDARCatv06} and \textit{G10CosmosCatv03}) with complete broadband coverage from $u$ to IRAC 2. We also remove AGN from this sample, as described by \citet{driver17} using the \citet{donley12} criteria (based on mid-infrared colours). This gives a final sample size of $\sim$149k objects in 0.915 deg$^2$. 

\section{The cosmic spectral energy distribution}
\label{sec:csed}

\subsection{Measuring the CSED}
\label{sec:csed2}

\begin{figure}
\begin{center}
\includegraphics[width=0.99\linewidth]{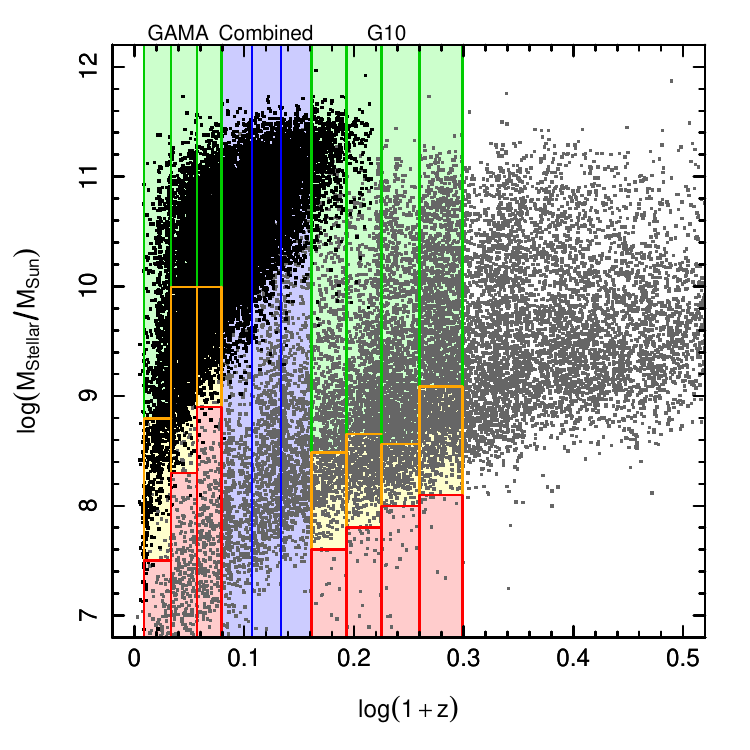}
\caption{Stellar mass versus redshift for the GAMA (black) and G10/COSMOS (grey) samples (1 in 10 plotted). Complete regions are denoted by green (as determined from the peak of the distribution of stellar masses in each redshift bin), incomplete regions are denoted by yellow and unobservable regions denoted by red. Redshift bins where a combined sample was used are denoted by blue.}
\label{fig:stellarmass}
\end{center}
\end{figure}

Figure \ref{fig:stellarmass} shows stellar mass as a function of redshift for both the GAMA (black) and G10/COSMOS (grey) samples. Stellar masses in both samples are derived from the \textsc{magphys} output. We define redshift bins of $0.02 < z < 0.08$, $0.08 < z < 0.14$ and $0.14 < z < 0.2$ for the GAMA data to be comparable with \citet{driver15} and $0.45 < z < 0.56$, $0.56 < z < 0.68$, $0.68 < z < 0.82$ and $0.82 < z < 0.99$ for G10/COSMOS. Each bin comprises of approximately 750 Myr in lookback time.

When using a flux-limited sample to estimate the resolved CSED, one inevitably misses objects that fall below either the apparent magnitude or surface brightness limits of the survey. As a consequence, the raw CSEDs derived from simply stacking the SEDs are not comparable across epochs. Firstly, each redshift bin requires a $1/V_\mathrm{max}$ correction to correct for incompleteness in the yellow shaded areas of Figure \ref{fig:stellarmass}. We compute these from the $r$ or $i^+$ absolute magnitude of each galaxy assuming limiting apparent magnitudes of $r = 19.8$ and $i^+ = 25.0$~mag for GAMA and G10 respectively. We then compute a (luminosity) distance and corresponding volume $V_\mathrm{max}$ for each object beyond which it should no longer be observable given these limits. Each object is then assigned a weight $w_i = (V_u - V_l) / (V_\mathrm{max} - V_l)$ where $V_u$ and $V_l$ correspond to the upper and lower edges of each redshift bin. These weights are capped at 10 to stop a single galaxy on the boundary of a particular redshift bin being upscaled to dominate emission in that bin. The (rest-frame) CSED $\epsilon(\lambda)$ can then be derived by simply computing:

\begin{equation}
\epsilon(\lambda) = \lambda L_\lambda = \frac{ \sum_i w_i \mathrm{SED}_i(\lambda)}{V_u - V_l}
\end{equation}

where SED$_i(\lambda)$ is the rest frame best fit SED of a galaxy in $\lambda f_\lambda$ units. This $V_\mathrm{max}$ correction biases the contribution to the CSED from lower mass systems by overweighting observed galaxies in a given redshift bin to compensate for systems below the apparent magnitude limit. This effect is mitigated by the small duration in lookback time (approximately 750 Myr each) of our redshift bins. Furthermore, in a sufficiently deep sample these systems only represent a small contribution to the total luminosity. This effect can only be addressed with deeper surveys. 

Secondly, each redshift bin samples a different range of stellar masses --- higher redshift bins are Malmquist biased towards high mass galaxies and will not include systems in the red shaded areas of Figure \ref{fig:stellarmass}. This incompleteness affects both the shape and normalization of the CSED because the CSED shape is mass dependent. 

\begin{figure*}
\begin{minipage}{7in}
\begin{center}
\includegraphics[width=0.99\linewidth]{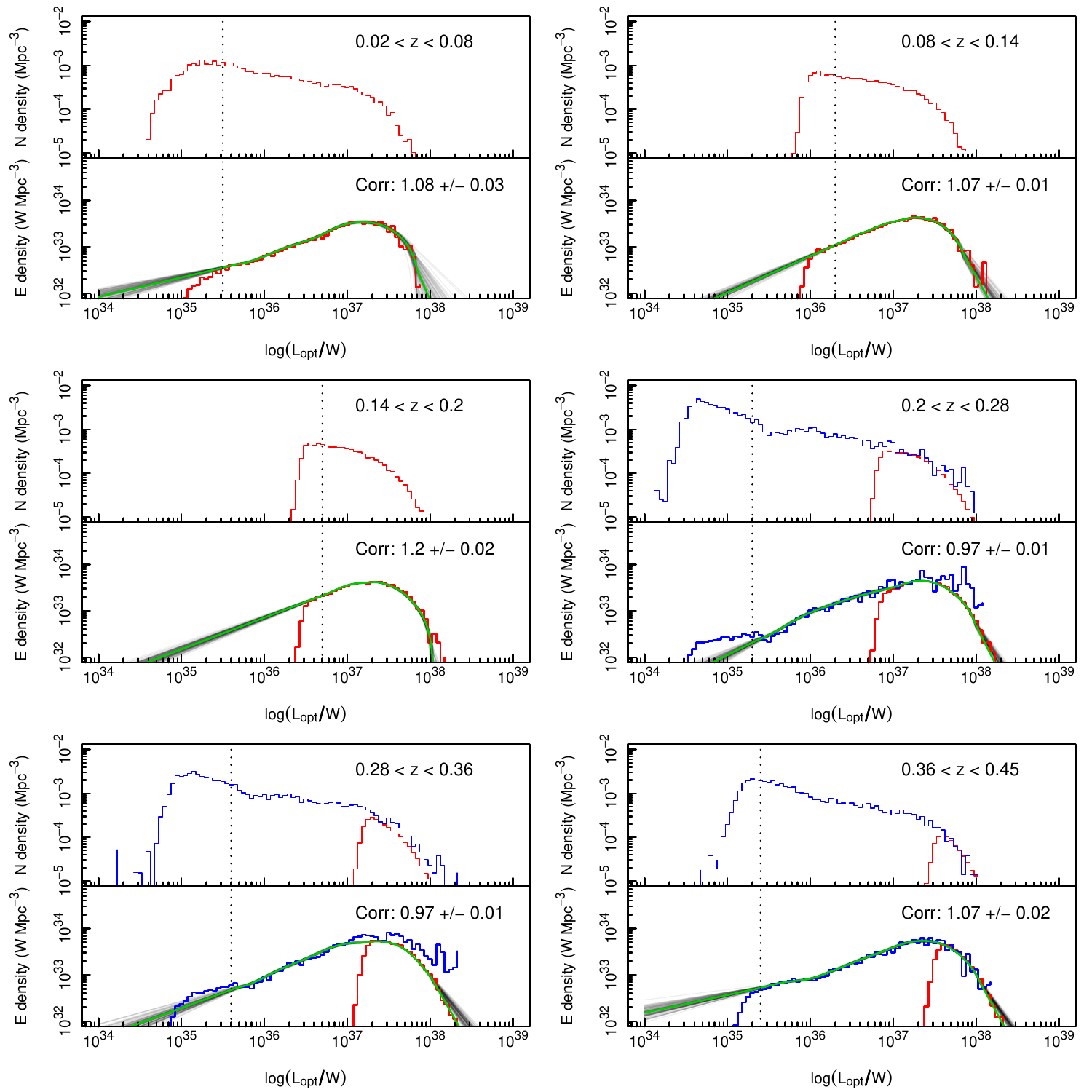}
\caption{Top panel: the total optical (100~nm $< \lambda < 8~\mu$m) luminosity function for each redshift interval in luminosity bins of log($L_\mathrm{opt}$) = 0.05 (GAMA: red, G10: blue). Bottom panel: the contribution to the total CSED of each luminosity bin (GAMA: red, G10: blue) fitted by a 10 point spline (green) to the completeness limit (black dashed vertical line). The correction to the CSED normalisation, computed from the ratio of integral of the spline to the histogram with error derived from 1000 Monte Carlo simulations (1 in 10 plotted), is also given.}
\label{fig:csedcorrect}
\end{center}
\end{minipage}
\end{figure*}

\begin{figure*}
\begin{minipage}{7in}
\begin{center}
\includegraphics[width=0.99\linewidth]{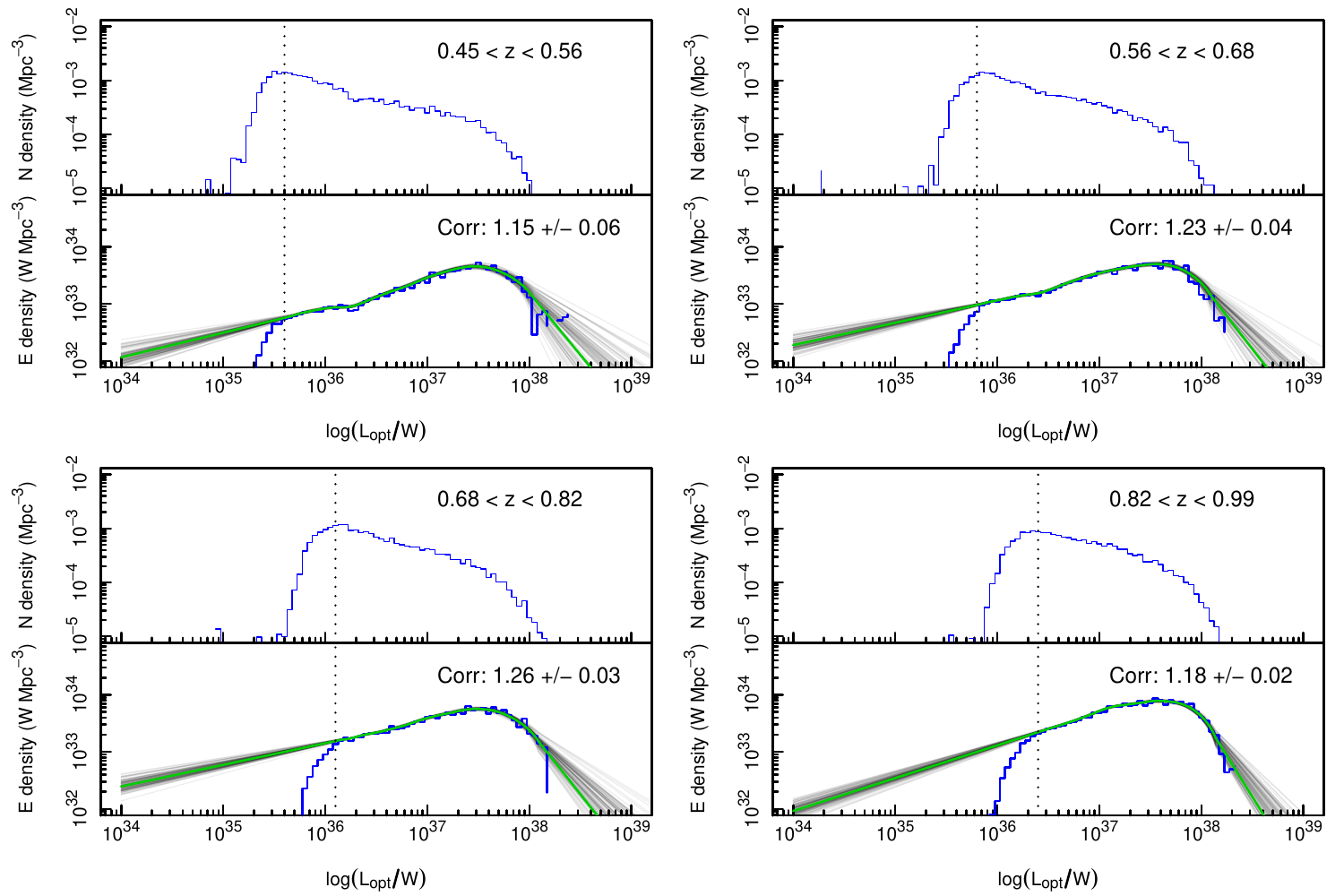}
\contcaption{}
\end{center}
\end{minipage}
\end{figure*}

To correct for the mass limit bias, we compute a total optical luminosity ($L_\mathrm{opt}$) for each galaxy by integrating the respective SED fit between 100~nm $ < \lambda < 8~\mu$m and construct a corresponding $L_\mathrm{opt}$ distribution (histogram) as shown in the top panels of Figure \ref{fig:csedcorrect}. We then compute the successive contributions of each bin to the CSED (bottom panel) by multiplying by $L_\mathrm{opt}$. Finally we fit a 10 point spline (green curve), weighted by the inverse fractional error squared, to where the energy density is complete and reliable. The fitting limit is determined by eye (see dotted lines on Figure \ref{fig:csedcorrect}). The spline fit is then extrapolated outside this range. Manual fitting helps ensure that the gradient of the extrapolated $L_\mathrm{opt}$ function is reasonable, reducing the impact of small number statistics at the high-mass end where the gradient is rapidly changing (see e.g. the $0.08 < z < 0.14$ and $0.45 < z < 0.56$ bins). The ratio of the integral under the spline to the total CSED (the integral under the red/blue lines) gives a redshift-independent correction factor as depicted in Figure \ref{fig:csedcorrect}. While the spline fit is bound at both ends, the integration is performed from $10^{34}$ to $10^{39}$~W (reflecting the extent of the data and to reduce extrapolation) to reduce the impact of error in the faint/bright end slope on the CSED. In the lowest redshift bin, extending the correction to $10^{33}$ W or $10^{30}$ W yields only 1.4 and 2.2 per cent extra flux respectively. The small bump in the $L_\mathrm{opt}$ distribution for the G10 in combined redshift bins is likely to be a systematic effect in either the photometric redshifts or SED fitting -- no such bump exists in the spectroscopic redshift only GAMA sample; we adjust the fitting range accordingly. Using a full bolometric luminosity (100~nm $ < \lambda < $ 1~mm) correction would be ideal for avoiding bias, however the quality of the far-infrared data does not permit this.

We also define three bins -- $0.20 < z < 0.28$, $0.28 < z < 0.36$ and $0.36 < z < 0.45$ -- where we combine the two samples (blue areas in Figure \ref{fig:stellarmass}). The ($V_\mathrm{max}$ corrected) samples are spliced such that the CSED is the sum of GAMA objects whose $L_\mathrm{opt} > 10^{37}$, $10^{37.2}$ and $10^{37.6}$~W (with increasing redshift) and G10 objects below these thresholds. The cutoff corresponds to the peak in the contribution to the total CSED of the GAMA sample. Both samples are consistent at high $L_\mathrm{opt}$, barring discontinuities due to cosmic variance. Note there is a significant overdensity in G10/COSMOS at $z \sim 0.35$ \citep{darvish17}. The combined sample is $L_\mathrm{opt}$ corrected in the manner described above.

\begin{figure*}
\begin{minipage}{7in}
\begin{center}
\includegraphics[width=0.99\linewidth]{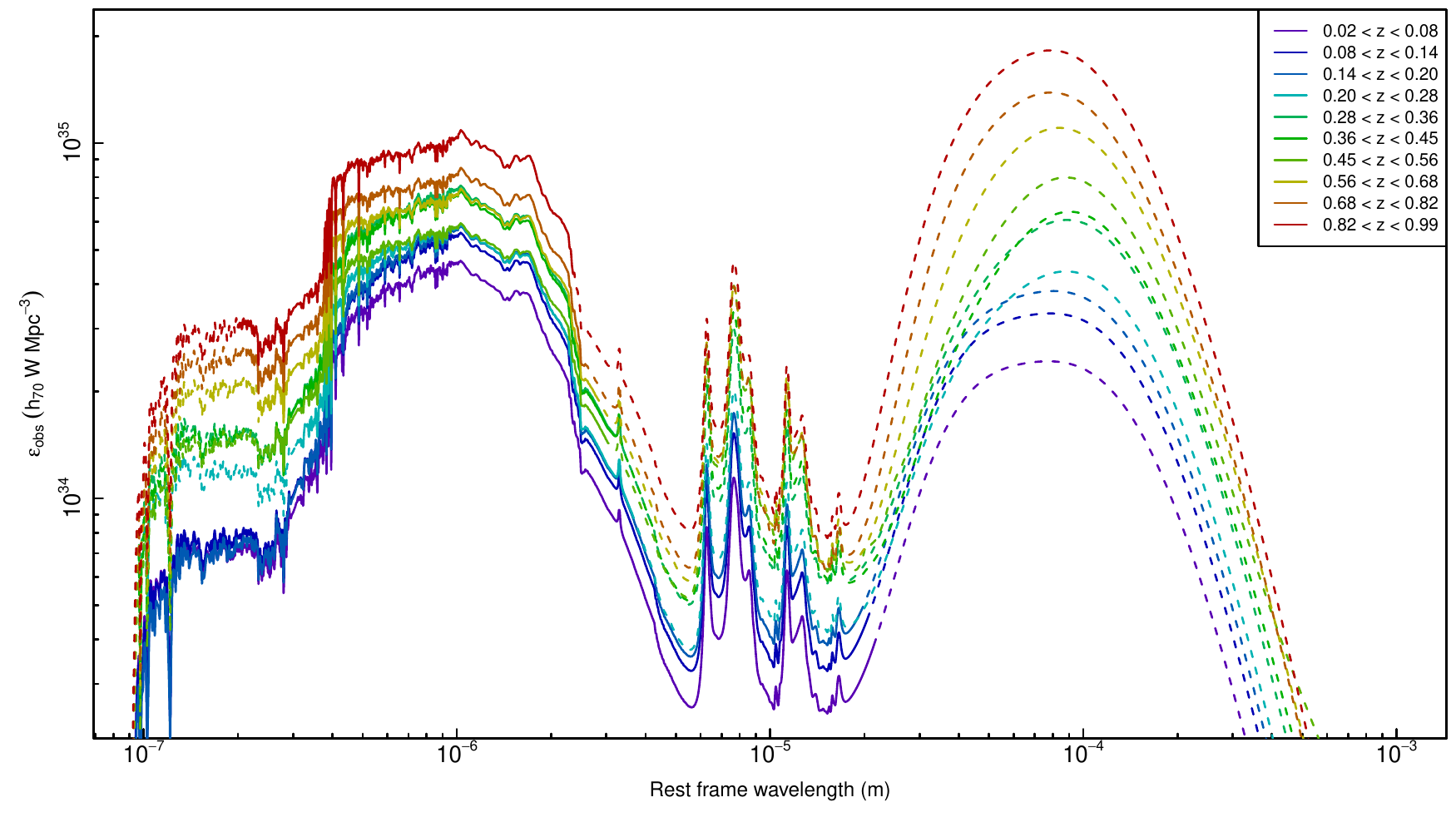}
\caption{The attenuated CSEDs for the GAMA and G10 datasets. Dashed lines indicate regions where the respective CSEDs are poorly constrained or partially extrapolated due to lack of data. The curves are subject to the normalisation errors described in Table \ref{tab:errors}, with uncertainty in the shape discussed in Section \ref{sec:csedbad}.}
\label{fig:acsed}
\end{center}
\end{minipage}
\end{figure*}

\begin{figure*}
\begin{minipage}{7in}
\begin{center}
\includegraphics[width=0.99\linewidth]{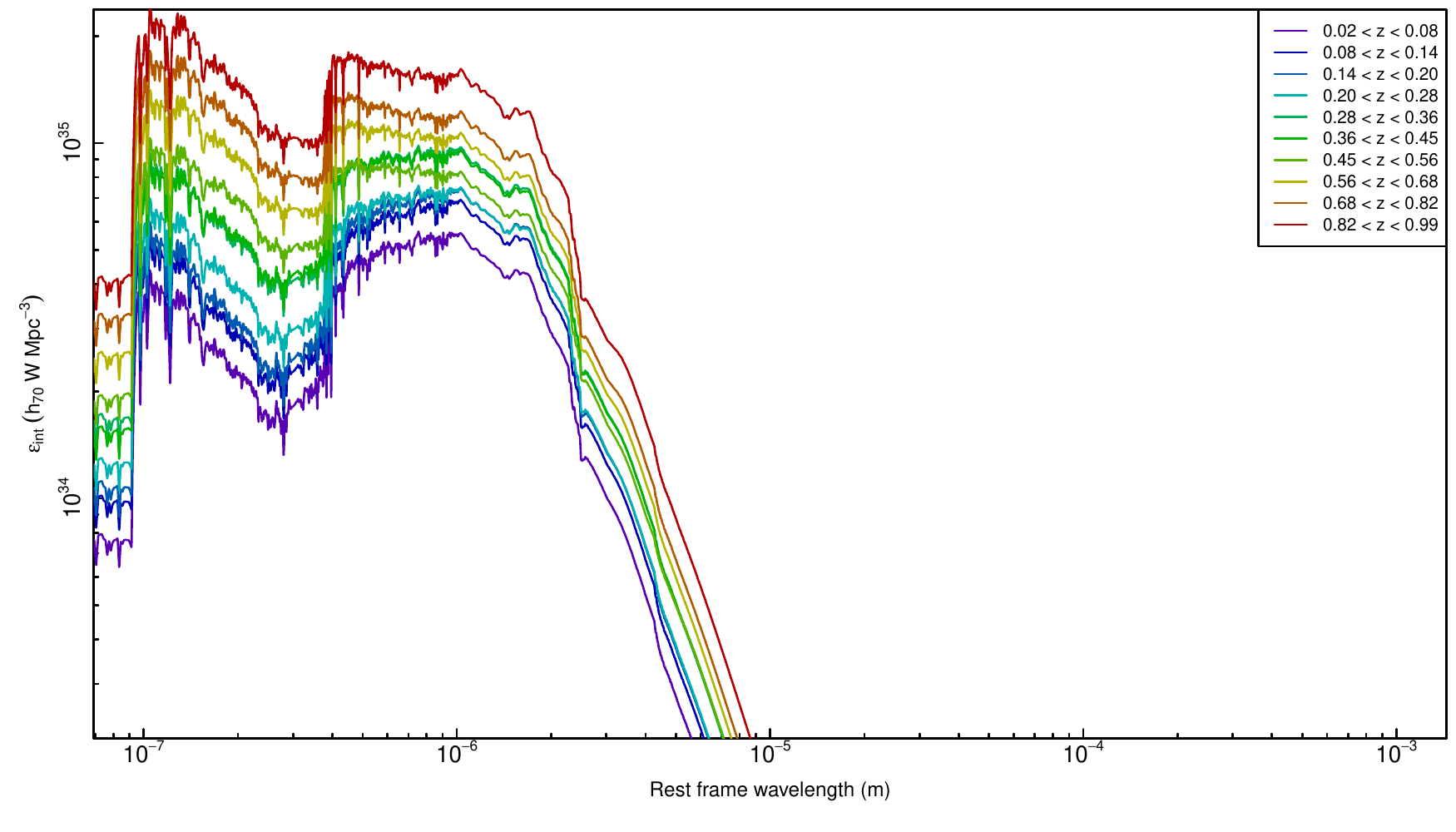}
\caption{The unattenuated CSEDs for the GAMA and G10 datasets. The curves are subject to the normalisation errors described in Table \ref{tab:errors}, with uncertainty in the shape discussed in Section \ref{sec:csedbad}.}
\label{fig:ucsed}
\end{center}
\end{minipage}
\end{figure*}

\begin{table*}
\begin{minipage}{7in}
\begin{center}
\caption{Attenuated, filter convolved CSED measurements for GAMA. At each epoch, three measurements are given -- (1) the sum of the $V_\mathrm{max}$ corrected fluxes, (2) the CSED convolved with the filter curve in the observed frame and (3) the sum of the $V_\mathrm{max}$ and $L_\mathrm{opt}$ corrected fluxes and upper limits. These measurements are subject to errors described in Table \ref{tab:errors}, photometric error is negligible.}
\label{tab:acsed_gama}
\begin{tabular}{l|c|ccc|ccc|ccc|}
\hline
Band	& Pivot      & \multicolumn{9}{c}{Attenuated, filter convolved CSED} \\
	& wavelength & \multicolumn{3}{c}{$z \in [0.02, 0.08]$} & \multicolumn{3}{c}{$z \in [0.08, 0.14]$} & \multicolumn{3}{c}{$z \in [0.14, 0.20]$} \\
	& ($\mu$m)   & \multicolumn{9}{c}{($10^{34}~h_{70}$~W~Mpc$^{-3}$)} \\
\multicolumn{2}{c}{Observed frame} & (1) & (2) & (3) & (1) & (2) & (3) & (1) & (2) & (3) \\
\hline
FUV 	  & 0.154 & 0.6 & 0.7 & 0.7 & 0.6 & 0.7 & 0.7 & 0.5 & 0.6 & 0.6 \\
NUV 	  & 0.230 & 0.7 & 0.7 & 0.7 & 0.8 & 0.8 & 0.8 & 0.6 & 0.7 & 0.8 \\
$u$ 	  & 0.356 & 1.1 & 1.1 & 1.2 & 1.1 & 1.1 & 1.2 & 0.9 & 1.0 & 1.0 \\
$g$	  & 0.470 & 2.5 & 2.8 & 2.7 & 2.6 & 2.8 & 2.8 & 2.0 & 2.4 & 2.4 \\
$r$ 	  & 0.616 & 3.4 & 3.8 & 3.7 & 3.9 & 4.2 & 4.1 & 3.3 & 4.1 & 4.0 \\
$i$  	  & 0.749 & 3.9 & 4.1 & 4.2 & 4.5 & 4.6 & 4.8 & 4.0 & 4.7 & 4.9 \\
$z$ 	  & 0.895 & 3.9 & 4.3 & 4.3 & 4.6 & 5.1 & 4.9 & 4.2 & 5.1 & 5.0 \\
$Z$ 	  & 0.880 & 4.0 & 4.3 & 4.3 & 4.8 & 5.1 & 5.1 & 4.3 & 5.1 & 5.2 \\
$Y$ 	  & 1.021 & 4.2 & 4.5 & 4.5 & 4.9 & 5.2 & 5.3 & 4.5 & 5.3 & 5.4 \\
$J$ 	  & 1.253 & 3.9 & 4.2 & 4.2 & 4.7 & 5.2 & 5.0 & 4.4 & 5.6 & 5.3 \\
$H$ 	  & 1.643 & 3.6 & 3.7 & 3.8 & 4.3 & 4.5 & 4.6 & 4.0 & 4.8 & 4.9 \\
$K$	  & 2.150 & 2.5 & 2.6 & 2.7 & 3.3 & 3.5 & 3.6 & 3.4 & 4.1 & 4.2 \\
W1	  & 3.370 & 0.8 & 0.9 & 0.9 & 1.1 & 1.2 & 1.2 & 1.1 & 1.4 & 1.3 \\
W2	  & 4.620 & 0.4 & 0.4 & 0.4 & 0.5 & 0.6 & 0.6 & 0.6 & 0.7 & 0.7 \\
W3	  & 12.10 & 0.3 & 0.4 & 0.4 & 0.5 & 0.6 & 0.5 & 0.5 & 0.7 & 0.6 \\
W4	  & 22.80 & 0.4 & 0.4 & 0.5 & 0.6 & 0.5 & 0.7 & 0.6 & 0.5 & 1.0 \\
PACS 100  & 101.0 & 2.0 & 2.3 & 2.7 & 2.8 & 3.2 & 4.0 & 3.1 & 3.8 & 5.3 \\
PACS 160  & 161.0 & 1.4 & 1.5 & 1.9 & 2.2 & 2.2 & 3.0 & 2.4 & 2.7 & 4.1 \\
SPIRE 250 & 249.0 & 0.5 & 0.6 & 0.6 & 0.8 & 1.0 & 0.9 & 0.9 & 1.3 & 1.2 \\
SPIRE 350 & 357.0 & 0.2 & 0.2 & 0.3 & 0.3 & 0.4 & 0.5 & 0.4 & 0.5 & 0.6 \\
SPIRE 500 & 504.0 & 0.1 & 0.1 & 0.1 & 0.1 & 0.1 & 0.3 & 0.2 & 0.1 & 0.4 \\
\hline
\end{tabular}
\end{center}
(1) Sum of $V_\mathrm{max}$ corrected fluxes. \\
(2) CSED convolved with the filter curve in the observed frame. \\
(3) Sum of $V_\mathrm{max}$ and $L_\mathrm{opt}$ corrected fluxes and upper limits.
\end{minipage}
\end{table*}

\begin{table*}
\begin{minipage}{7in}
\begin{center}
\caption{Attenuated, filter convolved CSED measurements for the intermediate redshift G10 region in the observed frame, as in Table \ref{tab:acsed_gama}. All measurements are subject to the errors described in Table \ref{tab:errors}.}
\label{tab:acsed_g10}
\begin{tabular}{l|c|ccc|ccc|ccc|ccc|}
\hline
Band	& Pivot      & \multicolumn{12}{c}{Attenuated, filter convolved CSED} \\
	& wavelength & \multicolumn{3}{c}{$z \in [0.45, 0.56]$} & \multicolumn{3}{c}{$z \in [0.56, 0.68]$} & \multicolumn{3}{c}{$z \in [0.68, 0.82]$} & \multicolumn{3}{c}{$z \in [0.82, 0.99]$} \\
        & ($\mu$m)   & \multicolumn{12}{c}{($10^{34}~h_{70}$~W~Mpc$^{-3}$)} \\
\multicolumn{2}{c}{Observed frame} & (1) & (2) & (3) & (1) & (2) & (3) & (1) & (2) & (3) & (1) & (2) & (3) \\
\hline
FUV 	  & 0.153 & 0.5 & 0.7 &  0.7 & 0.3 &  0.5 &  0.7 & 0.2 &  0.2 &  0.7 & 0.1 &  0.1 &  0.8 \\
NUV 	  & 0.225 & 0.9 & 1.4 &  1.4 & 1.1 &  1.8 &  1.9 & 1.2 &  2.0 &  2.2 & 1.3 &  2.1 &  2.5 \\
$u^*$ 	  & 0.381 & 1.2 & 1.4 &  1.3 & 1.5 &  1.9 &  1.8 & 1.8 &  2.5 &  2.3 & 2.4 &  3.0 &  2.9 \\
$g^+$	  & 0.478 & 1.4 & 1.8 &  1.6 & 1.7 &  2.2 &  2.1 & 1.9 &  2.5 &  2.4 & 2.6 &  2.8 &  3.0 \\
$r^+$ 	  & 0.630 & 3.0 & 3.7 &  3.4 & 3.1 &  3.9 &  3.8 & 3.0 &  3.5 &  3.8 & 3.2 &  3.6 &  3.8 \\
$i^+$  	  & 0.768 & 4.0 & 4.7 &  4.5 & 4.6 &  5.8 &  5.6 & 4.9 &  6.4 &  6.2 & 5.6 &  6.5 &  6.6 \\
$z^+$ 	  & 0.920 & 4.4 & 5.1 &  5.0 & 5.0 &  6.2 &  6.1 & 5.7 &  6.9 &  7.2 & 7.3 &  8.6 &  8.6 \\
$Y$ 	  & 1.021 & 4.4 & 5.2 &  5.0 & 5.1 &  6.4 &  6.1 & 5.4 &  7.3 &  6.8 & 7.0 &  8.8 &  8.3 \\
$J$ 	  & 1.252 & 4.4 & 5.5 &  5.0 & 5.1 &  6.7 &  6.3 & 5.9 &  7.5 &  7.4 & 8.0 &  9.2 &  9.5 \\
$H$ 	  & 1.643 & 4.3 & 5.6 &  4.9 & 5.1 &  7.1 &  6.3 & 5.8 &  7.9 &  7.3 & 7.7 &  9.7 &  9.1 \\
$K$	  & 2.150 & 4.0 & 4.9 &  4.6 & 5.1 &  6.3 &  6.2 & 5.8 &  7.6 &  7.3 & 8.1 & 10.1 &  9.6 \\
IRAC 1	  & 3.560 & 1.8 & 2.5 &  2.1 & 2.8 &  3.8 &  3.4 & 3.7 &  5.1 &  4.6 & 6.2 &  7.6 &  7.3 \\
IRAC 2	  & 4.510 & 1.1 & 1.5 &  1.3 & 1.6 &  2.0 &  1.9 & 2.0 &  2.6 &  2.5 & 3.6 &  4.5 &  4.3 \\
IRAC 3	  & 5.760 & 0.6 & 1.0 &  1.0 & 0.9 &  1.4 &  1.7 & 1.1 &  1.8 &  2.1 & 1.8 &  2.5 &  3.0 \\
IRAC 4	  & 7.960 & 0.5 & 0.7 &  0.9 & 0.5 &  0.7 &  1.1 & 0.6 &  0.9 &  1.3 & 1.0 &  1.4 &  1.9 \\
MIPS 24   & 23.68 & 0.5 & 0.7 &  1.0 & 0.5 &  0.8 &  1.5 & 0.6 &  1.0 &  1.9 & 1.0 &  1.5 &  2.7 \\
MIPS 70	  & 71.42 & 0.8 & 4.7 & 16.8 & 0.7 &  6.2 & 27.9 & 0.9 &  7.8 & 36.6 & 1.3 &  9.4 & 47.1 \\
PACS 100  & 101.0 & 3.9 & 7.1 & 26.0 & 5.1 &  9.7 & 42.8 & 5.8 & 12.3 & 55.7 & 8.5 & 15.4 & 71.3 \\
PACS 160  & 161.0 & 3.8 & 7.2 & 28.2 & 5.2 & 10.1 & 47.0 & 6.3 & 13.0 & 61.6 & 9.5 & 17.5 & 79.3 \\
SPIRE 250 & 249.0 & 2.1 & 4.4 &  9.7 & 3.1 &  6.4 & 16.3 & 3.8 &  8.6 & 21.3 & 6.2 & 12.7 & 28.2 \\
SPIRE 350 & 357.0 & 0.7 & 2.1 &  7.8 & 1.2 &  3.2 & 13.3 & 1.4 &  4.3 & 17.4 & 2.6 &  6.8 & 22.9 \\
SPIRE 500 & 504.0 & 0.2 & 0.8 &  6.6 & 0.3 &  1.2 & 11.2 & 0.4 &  1.6 & 14.7 & 0.7 &  2.6 & 19.1 \\
\hline
\end{tabular}
\end{center}
(1) Sum of $V_\mathrm{max}$ corrected fluxes. \\
(2) CSED convolved with the filter curve in the observed frame. \\
(3) Sum of $V_\mathrm{max}$ and $L_\mathrm{opt}$ corrected fluxes and upper limits.
\end{minipage}
\end{table*}

\begin{table*}
\begin{minipage}{7in}
\begin{center}
\caption{Unattenuated, filter convolved CSED measurements for the given filters for GAMA in the observed frame. The unattenuated CSED is negligible in bands longwards of W2. All estimates are subject to the errors described in Table \ref{tab:errors}.}
\label{tab:ucsed}
\begin{tabular}{l|c|ccc|}
\hline
Band	& Pivot      & \multicolumn{3}{c}{Unattenuated, filter convolved CSED} \\
	& wavelength & $z \in [0.02, 0.08]$ & $z \in [0.08, 0.14]$ & $z \in [0.14, 0.20]$  \\
	& ($\mu$m)   & \multicolumn{3}{c}{($10^{34}~h_{70}$~W~Mpc$^{-3}$)} \\
\multicolumn{2}{c}{Observed frame} & \multicolumn{3}{c}{} \\
\hline
FUV 	& 0.154 & 3.1 & 4.1 & 4.6 \\
NUV 	& 0.230 & 2.0 & 2.7 & 3.1 \\
$u$ 	& 0.356 & 2.0 & 2.4 & 2.4 \\
$g$	& 0.470 & 4.0 & 4.5 & 4.2 \\
$r$ 	& 0.618 & 5.0 & 5.9 & 6.2 \\
$i$  	& 0.749 & 5.2 & 6.2 & 6.7 \\
$z$ 	& 0.895 & 5.3 & 6.6 & 7.0 \\
$Z$     & 0.880 & 5.3 & 6.6 & 7.0 \\
$Y$ 	& 1.021 & 5.5 & 6.6 & 7.0 \\
$J$ 	& 1.252 & 4.9 & 6.3 & 7.1 \\
$H$ 	& 1.643 & 4.3 & 5.3 & 5.8 \\
$K$	& 2.150 & 2.9 & 4.0 & 4.9 \\
W1      & 3.370 & 0.9 & 1.3 & 1.5 \\
W2      & 4.620 & 0.4 & 0.6 & 0.8 \\
\hline
\end{tabular}
\end{center}
\end{minipage}
\end{table*}

\begin{table*}
\begin{minipage}{7in}
\begin{center}
\caption{Unattenuated, filter convolved CSED measurements convolved for given filters for G10 in the observed frame, as in Table \ref{tab:ucsed}. The unattenuated CSED is negligible in bands longwards of W2. All estimates are subject to the errors described in Table \ref{tab:errors}.}
\label{tab:ucsed_g10}
\begin{tabular}{l|c|cccc|}
\hline
Band	& Pivot      & \multicolumn{4}{c}{Unattenuated, filter convolved CSED} \\
	& wavelength & $z \in [0.45, 0.56]$ & $z \in [0.56, 0.68]$ & $z \in [0.68, 0.82]$ & $z \in [0.82, 0.99]$ \\
	& ($\mu$m)   & \multicolumn{4}{c}{($10^{34}~h_{70}$~W~Mpc$^{-3}$)} \\ 
\multicolumn{2}{c}{Observed frame} & \multicolumn{4}{c}{} \\
\hline
FUV 	& 0.154 &  7.6 &  7.0 &  5.1 &  4.1 \\
NUV 	& 0.230 &  8.1 & 11.4 & 15.0 & 19.8 \\
$u^*$ 	& 0.381 &  5.3 &  7.3 & 10.3 & 14.5 \\
$g^+$	& 0.478 &  5.1 &  6.5 &  8.2 & 11.1 \\
$r^+$ 	& 0.630 &  7.6 &  8.5 &  8.5 & 10.1 \\
$i^+$  	& 0.768 &  8.5 & 10.9 & 12.9 & 14.5 \\
$z^+$ 	& 0.920 &  8.5 & 10.8 & 12.7 & 17.2 \\
$Y$ 	& 1.021 &  8.3 & 10.7 & 12.6 & 16.4 \\
$J$ 	& 1.252 &  8.2 & 10.4 & 12.0 & 15.7 \\
$H$ 	& 1.643 &  7.8 & 10.3 & 11.7 & 15.1 \\
$K$	& 2.150 &  6.4 &  8.7 & 10.6 & 14.7 \\
IRAC 1  & 3.370 &  3.0 &  4.8 &  6.4 & 10.0 \\
IRAC 2  & 4.620 &  1.7 &  2.3 &  3.1 &  5.6 \\
IRAC 3  & 5.760 &  1.0 &  1.5 &  2.0 &  2.9 \\
IRAC 4  & 7.960 &  0.5 &  0.7 &  0.9 &  1.5 \\
\hline
\end{tabular}
\end{center}
\end{minipage}
\end{table*}

We show the resulting attenuated and unattenuated CSEDs, rescaled using the correction factors shown in Figure \ref{fig:csedcorrect}, in Figures \ref{fig:acsed} and \ref{fig:ucsed} and Tables \ref{tab:acsed_gama}, \ref{tab:acsed_g10}, \ref{tab:ucsed} and \ref{tab:ucsed_g10}. The dotted lines in the Figures show regions where CSEDs are potentially unreliable due to the underlying data either being missing or of too low sensitivity and resolution as described in \citet{driver17}.

\subsection{The CSED error budget}
\label{sec:csederr}

\begin{table}
\begin{center}
\caption{The CSED error terms arising from cosmic variance (CV), jackknife resampling, use of photometric redshifts and the $L_\mathrm{opt}$ correction. The total error in per cent for a given redshift bin is the sum of all columns in quadrature for that bin.}
\label{tab:errors}
\begin{tabular}{l|ccccc|}
\hline
Redshift      & CV & Jackknife & Photo-z & $L_\mathrm{opt}$ & Total \\
& (\%) & (\%) & (\%) & (\%) & (\%) \\
\hline
$0.02 - 0.08$ & 20 & 3         & 0       & 3	& 20 \\
$0.08 - 0.14$ & 14 & 1         & 0       & 1	& 14 \\
$0.14 - 0.20$ & 11 & 1         & 0       & 2	& 11 \\
$0.20 - 0.28$ & 22 & 2         & 2       & 1    & 22 \\
$0.28 - 0.36$ & 25 & 2         & 3       & 1    & 25 \\
$0.36 - 0.45$ & 27 & 2         & 4       & 2    & 27 \\
$0.45 - 0.56$ & 26 & 4         & 5       & 5	& 27 \\
$0.56 - 0.68$ & 24 & 4         & 5       & 4 	& 25 \\
$0.68 - 0.82$ & 21 & 3         & 5       & 3	& 22 \\
$0.82 - 0.99$ & 18 & 1         & 5       & 2	& 19 \\
\hline
\end{tabular}
\end{center}
\end{table}

Using the prescription in \citet{driver10}\footnote{\url{http://cosmocalc.icrar.org}}, we derive the cosmic (sample) variance (CV) for each of our redshift bins based on areas of 129 deg$^2$ for GAMA and 0.915 deg$^2$ for the G10 region (see Table \ref{tab:errors}). For the combined bins, we compute the fraction of $L_\mathrm{opt}$ (55, 37 and 16 per cent for GAMA) arising from each of GAMA and G10/COSMOS;  CV is the weighted average (by $L_\mathrm{opt}$) of the respective CVs of GAMA and G10/COSMOS. The \citet{driver12} CSED is subject to 5 per cent CV (derived by bootstrapping to the Sloan Digital Sky Survey and using the Driver \& Robotham recipe) and approximately 5 per cent error resulting from the luminosity function fits. The \citet{driver15} CSEDs have 18, 12 and 10 per cent CV in order of increasing redshift with completeness and photometric errors not estimated. 

The majority of redshifts in the G10/COSMOS sample are extracted from the \citet{laigle16} photometric redshift catalogue -- 22, 19, 18 and 15 per cent of objects have spectroscopic redshifts in the defined redshift bins respectively. \citet{laigle16} claim a normalized median absolute deviation of 0.007 against a sample of zCOSMOS-bright spectroscopic redshifts. However, to be conservative, we add an extra 5 per cent to the G10/COSMOS CSED error budget in quadrature to account for unknown systematics.

We also perform a jackknife resampling by deleting 10 per cent of the sample (as determined by the last digit of the GAMA/G10 catalogue number) for 10 iterations to check whether any given CSED is being dominated by a small number of SED fits. The error is given by $\sigma^2 = \frac{N-1}{N} \sum_{i=1}^N mean(x_i^2 - x^2)$, where N = number of iterations, $mean(f)$ is the mean of $f$ across 100~nm~$< \lambda < 24~\mu$m (to avoid regions where the CSED is substantially extrapolated), $x_i$ is the $i$th resampled CSED and $x$ the total CSED. This analysis shows our CSED stacks are robust, with errors not exceeding 2 per cent.

The $L_\mathrm{opt}$ correction (Figure \ref{fig:csedcorrect}) is also subject to uncertainty. We estimate this error in 1000 Monte Carlo simulations. In each simulation, each bin of the contribution to the CSED is perturbed by a normal distribution with mean zero and standard deviation equal to the Poissonian error. We then refit the spline and rederive the correction factors. The quoted error ($L_\mathrm{opt}$ in Table \ref{tab:errors}) is the standard deviation of correction factors. 

The final error budget is shown in Table \ref{tab:errors}, with the entries combined in quadrature. As expected, CV is the dominant error term. However, all error terms will be reduced by next-generation spectroscopic surveys of galaxy evolution, such as the Wide Area VISTA Extragalactic Survey (WAVES; \citealt{waves}) -- which will probe significantly larger areas to higher redshifts. 

\subsection{Biases, caveats and missing energy}
\label{sec:csedbad}

\begin{figure*}
\begin{minipage}{7in}
\begin{center}
\includegraphics[width=0.99\linewidth]{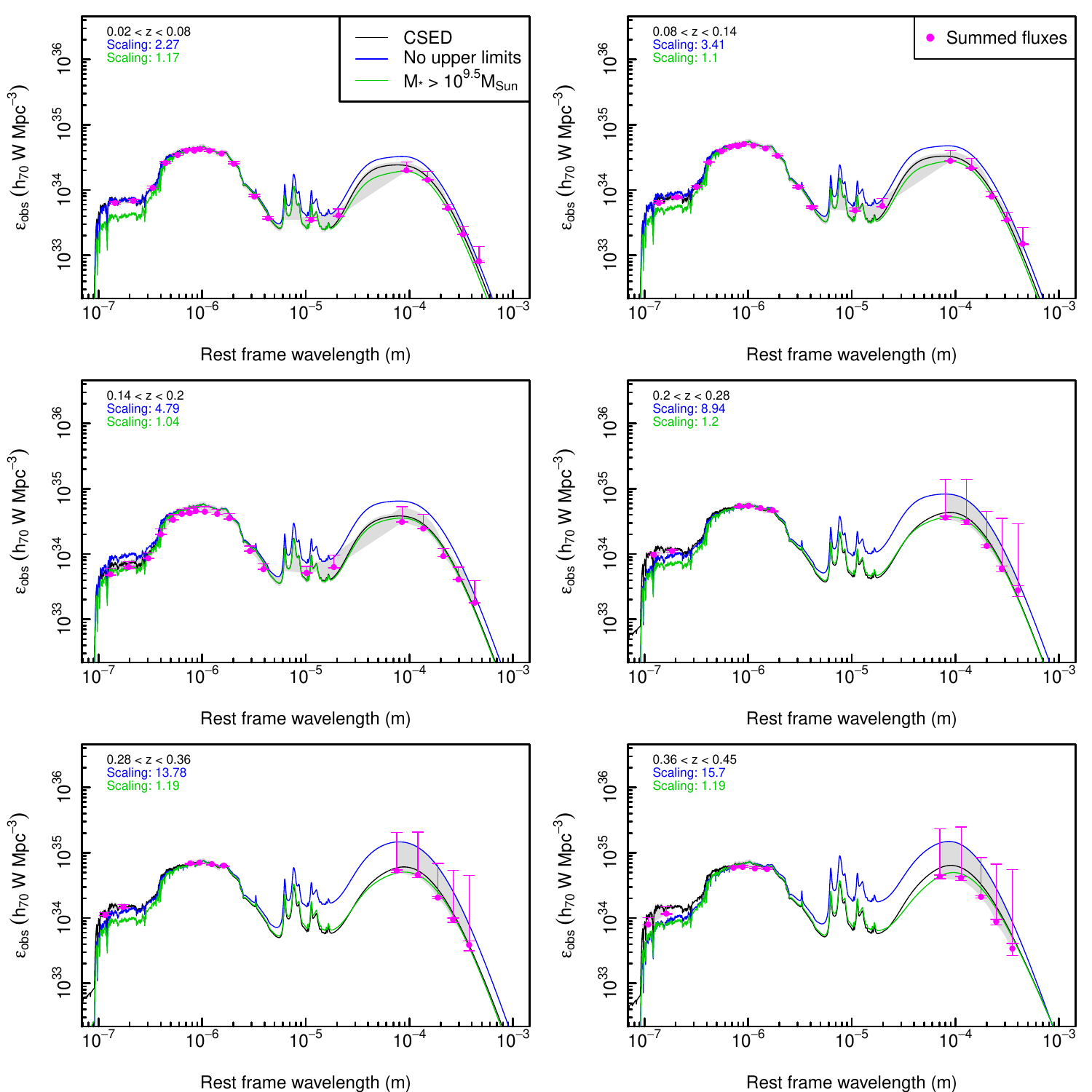}
\caption{CSEDs comprising all galaxies in a redshift bin (black), those with $M_\star > 10^{9.5} M_\odot$ (green) and those with complete photometric measurements only (i.e. no upper limits, blue). All CSEDs are normalized at 1~$\mu$m with the scaling factor given. The $V_\mathrm{max}$ corrected sum of fluxes with errors summed in quadrature is also shown (pink points), as is the $V_\mathrm{max}$ and $L_\mathrm{opt}$ corrected sum of fluxes and upper limits (upper pink error bar). The $V_\mathrm{max}$ corrected sum of fluxes with errors from objects excluded as AGN is given in orange points for G10 only. Negligible errors are not plotted. The grey and light orange areas are indicative estimates of the SED modelling error and error in AGN CSED respectively. For bins with combined GAMA and G10/COSMOS data, we show only the common filters.}
\label{fig:csedbad}
\end{center}
\end{minipage}
\end{figure*}

\begin{figure*}
\begin{minipage}{7in}
\begin{center}
\includegraphics[width=0.99\linewidth]{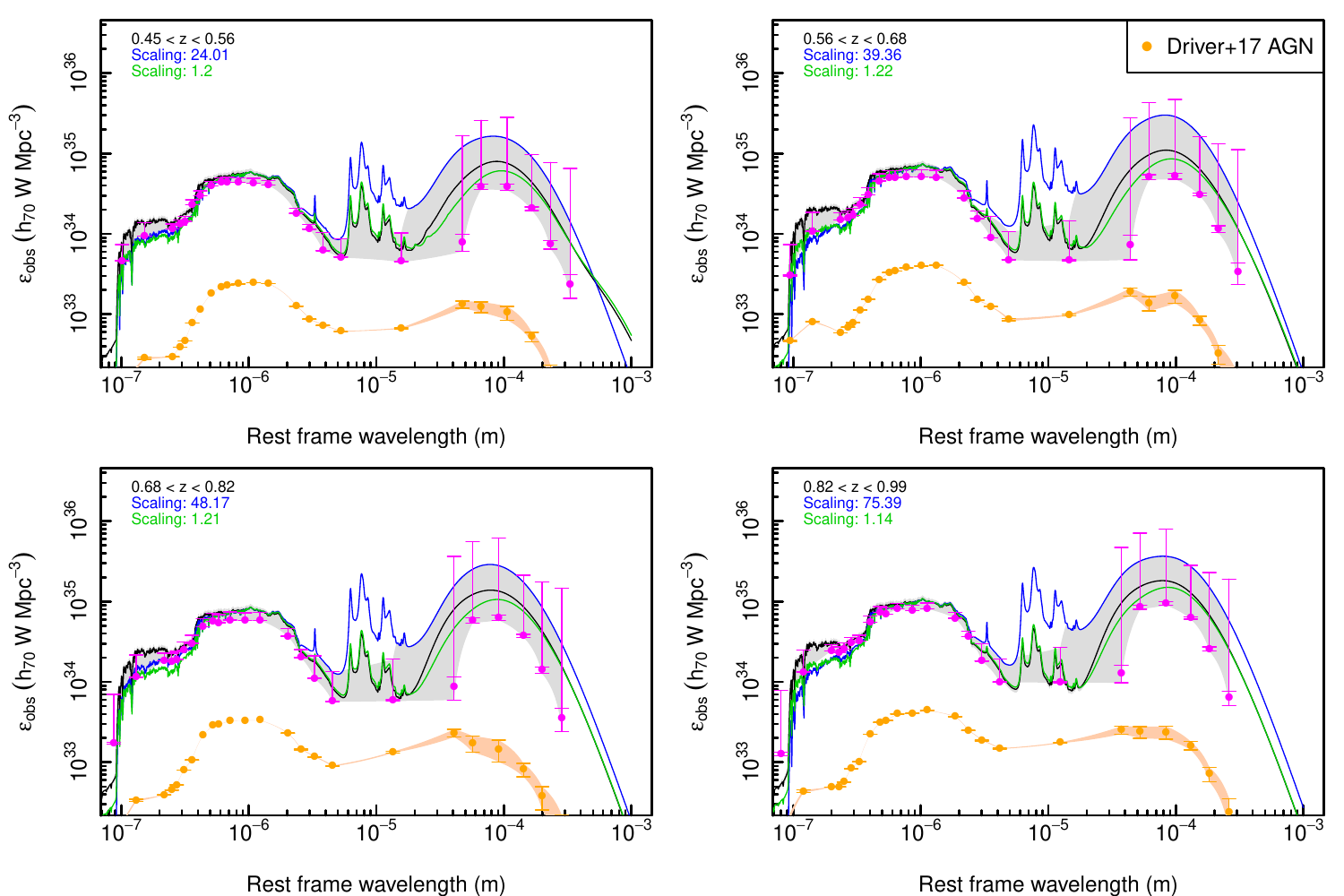}
\contcaption{}
\end{center}
\end{minipage}
\end{figure*}

\begin{figure}
\begin{center}
\includegraphics[width=0.99\linewidth]{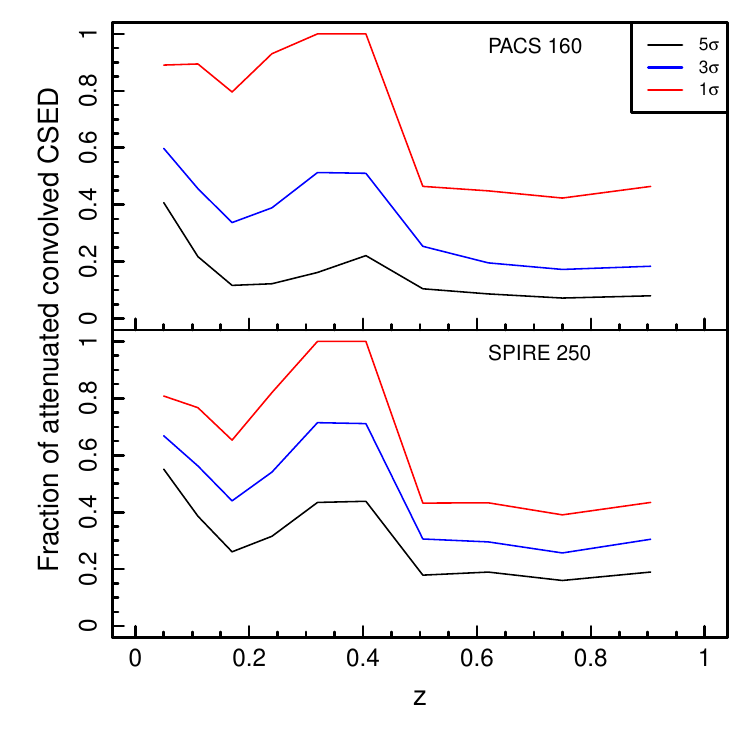}
\caption{The fraction of the attenuated CSED convolved with the given filters in the observed frame accounted for by summing $V_\mathrm{max}$ (but not $L_\mathrm{opt}$) corrected fluxes with the given significance.}
\label{fig:firextrapolation}
\end{center}
\end{figure}

\textsc{magphys} does not provide an error estimate on the interpolated (best fit) spectrum. We create an indicative estimate of the uncertainty in the CSED shape by recomputing the CSED within each volume through four distinct methods:

\begin{enumerate}
\item A strict lower bound to the CSED can be derived by summing the $V_\mathrm{max}$ corrected catalogue's fluxes and errors (in quadrature) for each galaxy in each redshift bin (i.e. ignoring the $L_\mathrm{opt}$ correction). Non-detections and non-measurements are treated as zero flux with an error equal to the appropriate $1\sigma$ upper limit. This is consistent with the use of $1\sigma$ limits in both the underlying photometry and SED fitting. These values are shown in Tables \ref{tab:acsed_gama} and \ref{tab:acsed_g10} (first in each set of three measurements) and Figure \ref{fig:csedbad} (pink points). As expected, photometric errors scale with resolution and sensitivity, with the high-resolution optical and near-infrared being associated with negligible error and the low-sensitivity far-infrared data in G10 having the largest error.

\item A strict upper bound can be derived from the sum of the $V_\mathrm{max}$ and $L_\mathrm{opt}$ corrected fluxes and treating non-detections and non-measurements as having flux equal to the $1\sigma$ upper limit. We present this in Tables \ref{tab:acsed_gama} and \ref{tab:acsed_g10} (third in each set of three measurements) and as the upper pink error bar in Figure \ref{fig:csedbad}.

\item We compute the CSED for galaxies with non-zero flux in all photometric bands in each redshift bin and rescale to the full CSED at $1~\mu$m. We depict these CSEDs and respective normalisation constants with blue curves on Figure \ref{fig:csedbad}. This sample is biased toward dusty systems, especially those with colder dust populations (due to relatively low sensitivity at 500~$\mu$m), and thus provides an alternative worst case scenario (upper limit) for the far-infrared CSED. This curve does not have the same properties in the ultraviolet part of the spectrum because some sources with measurable flux are excluded. 

\item We compute the CSED for galaxies with log$(M_*) > 9.5$ in each redshift bin and renormalise to the full CSED at $1~\mu$m. We depict these CSEDs and respective normalisation constants with green curves on Figure \ref{fig:csedbad}. This sample is biased toward massive, quiescent, dust free galaxies. As our redshift bins are complete to log$(M_*) > 9.5$, the mass-restricted CSED represents a lower limit to the ultraviolet CSED.
\end{enumerate}

We then combine these alternative estimates of the CSED to obtain an indicative error range. For the upper bound, we take the minimum of the rescaled no upper limits CSED (blue curve) and a spline interpolation of the sum of fluxes and upper limits (upper pink error bar). The lower bound is a spline interpolation of the photometric measurements minus the photometric error. We also impose a minimum 10 per cent error from the measured CSED (black curve) to account for SED modelling error. The result is the grey regions in Figure \ref{fig:csedbad}. The error terms described in Section \ref{sec:csederr} should be added to this range in quadrature. 

The large errors present in the far-infrared portion of G10/COSMOS demonstrate the CSEDs in that region are mostly extrapolated and are depicted with dotted lines in Figure \ref{fig:acsed}. Similarly, we depict the GAMA CSEDs with dotted lines beyond $23~\mu$m; the underlying lack of $70~\mu$m imaging prevents the CSED being constrained between 23 and 100 $\mu$m. The relatively low detection rate in the \textit{Herschel} bands reduces the reliability of the far-infrared CSED. Figure \ref{fig:firextrapolation} shows the fraction of the attenuated, filter convolved (in the observed frame) CSED accounted for by summing $V_\mathrm{max}$ (but not $L_\mathrm{opt}$) corrected fluxes in PACS 160 and SPIRE 250. 20 -- 70 per cent of the attenuated, filter convoved CSED is accounted for by $>3\sigma$ detections depending on redshift. The majority of the CSED at 160 and 250 $\mu$m below $z = 0.45$ is accounted for by $>1\sigma$ excesses over the background, while above this redshift this fraction is about half. The remainder consists of interpolated and extrapolated flux.

To estimate the effect of incompleteness on the CSED shape, we refer to the ratio of ultraviolet emission between the mass-restricted (green) and unrestricted (black) CSEDs. Assuming that this ratio does not change with redshift, this suggests that the higher redshift GAMA bins are potentially missing significant ($\sim$20 per cent) ultraviolet flux. We do not correct for this effect because it is likely to be redshift dependent. Some of this emission will be attenuated by dust and appear in the far-infrared. This should not significantly affect the overall energy output due to the relatively low amount of energy escaping into the intergalactic medium at these wavelengths.

We also consider the effect of AGN contamination on the CSED. \textsc{magphys} does not incorporate AGN emission into its template library. In G10, we sum fluxes for the objects excluded from the \citet{driver17} analysis as AGN using the \citet{donley12} criteria. Summing fluxes and errors in quadrature for these objects gives the orange points in Figure \ref{fig:csedbad}. When compared to the sum of fluxes for all objects that enter the CSED sample, AGN and their host galaxies consist of $\sim$4 per cent of the (observed frame) optical and near-infrared flux but $\sim10$ per cent in the mid-infrared for $z = 0.505$ and 12-17 per cent for $z = 0.905$. As an aside, it is interesting to note that AGN at these redshifts appear to be entirely dominated by their quiescent host galaxies and/or are heavily obscured. For GAMA and the combined bins, we assume AGN contamination to be negligible due to low number density at $z < 0.5$ (in line with \citealt{driver17}). 

At ultraviolet wavelengths, the \citet{driver12} and the \citet{driver15} CSEDs are based on curve of growth photometry that captures emission from extended UV discs \citep{gildepaz05,thilker07}. The \citet{wright16a} catalogue computes consistently derived aperture-matched photometry (defined in the $r$ band) across all wavelengths. This approach trades potentially missing extended flux for consistent errors from the ultraviolet to the infrared, with the latter being more advantageous for SED fitting.

We do not see a significant impact on the near-infrared CSED at low redshifts caused by uncertainty in modelling thermally-pulsating asymptotic giant branch (TP-AGB) stars (see e.g. \citealt{maraston06,bruzual07,conroy10,bruzual13,noel13,capozzi16}) -- our CSEDs lie within the photometric bounds for GAMA. It is unknown whether \textsc{magphys} compensates for imprecision in TP-AGB modelling by adjusting the other SED fitting parameters (such as stellar phase metallicity or ages) in order to achieve a mathematically good, but unphysical, fit. Using a SED fitting code based on the \citet{maraston05} (which arguably overestimate the TP-AGB contribution) or other stellar library to determine the uncertainty range would naturally be part of a thorough evaluation of SED fitting, which is also out of scope of this paper.

The aperture photometry used within may systematically miss flux in highly concentrated systems, which are typically massive, quiescent and dust-free galaxies. One could determine an aperture correction by fitting a S\'{e}rsic profile to each source and computing the ratio of the integrated flux to the aperture flux (see e.g. \citealt{taylor11}). These come with their own set of biases and random errors, which may be wavelength dependent. No such analysis has been performed on G10/COSMOS, so for consistency across the redshift range we do not include it.

The $L_\mathrm{opt}$ correction only serves to adjust the normalisation of the CSED. This creates a bias against fainter systems if they have substantially different SEDs than the average population and, in aggregate, contribute significantly to the total CSED. This may be evident in e.g. the peak of the far-infrared CSED at $0.14 < z < 0.2$ compared to $0.2 < z < 0.28$, with the caveat that a greater portion of the far-infrared CSED arises from extrapolation in the higher redshift bin and the complete lack of 70~$\mu$m data in GAMA compared to at least an upper limit in G10/COSMOS. This is best addressed with deeper data.

The $r$ and $i^+$ selections used to define the GAMA and G10/COSMOS samples are equivalent to a luminosity cut at successively shorter wavelengths in the rest frame with redshift. The most obvious resulting bias is against highly obscured or quiescent systems which lie below the respective magnitude limits. More subtle biases may occur depending on the SEDs of individual sub-populations. The largest effect occurs where $r$ or $i^+$ is equivalent to $u$ (and to a lesser extent, $g$) in the rest frame, namely the high-redshift ends of the GAMA and G10/COSMOS samples. This is not an issue in GAMA out to $z = 0.10$, as our low-redshift CSED is consistent with that constructed from luminosity functions in each band from FUV through $K$ by \citet{driver12}. The effect of these biases can only be quantified with a deep, near-infrared selected sample, which will also enable studies of the CSED at higher redshifts.

\subsection{The energy output of the Universe}
\begin{figure}
\begin{center}
\includegraphics[width=0.99\linewidth]{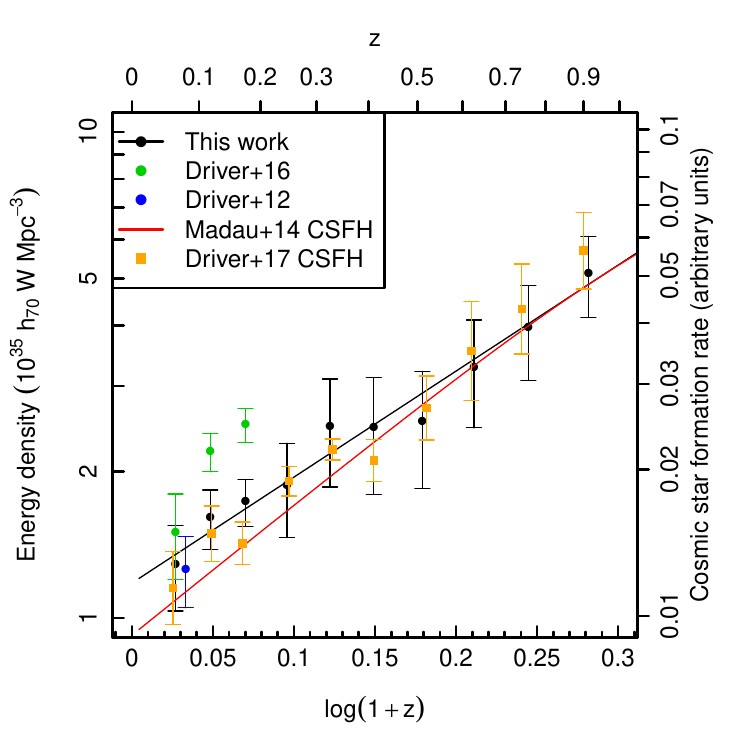}
\caption{The energy output of the Universe as measured from this work, \citet{driver12} and \citet{driver15}. Errors depicted are those described in Table \ref{tab:errors} only. For comparison, we add the \citet{driver17} cosmic star formation rate density measurements and the \citet{madau14} cosmic star formation rate density fitting function, arbitrarily scaled to the fit to the total energy output for the highest redshift bin.}
\label{fig:summary}
\end{center}
\end{figure}

\begin{figure*}
\begin{minipage}{7in}
\begin{center}
\includegraphics[width=0.99\linewidth]{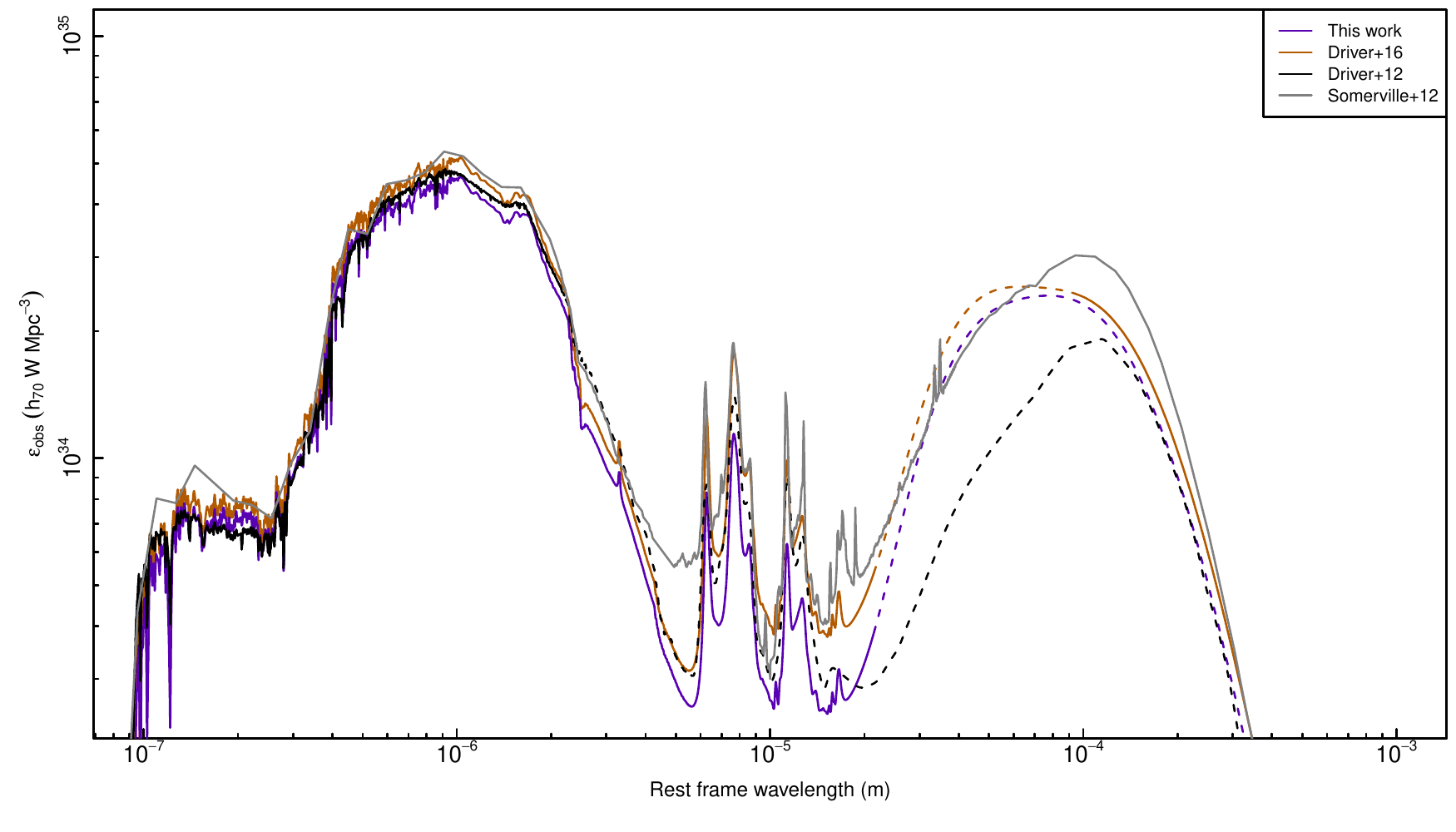}
\caption{The attenuated CSEDs for the GAMA dataset compared to previous estimates for $0.02 < z < 0.08$. The grey, black and orange lines depict the \citet{somerville12} semi-analytic model, the \citet{driver12} data and the \citet{driver15} respectively. Dashed lines indicate regions where the respective CSEDs are poorly constrained due to lack of data. The curves are subject to the errors described in Table \ref{tab:errors}.}
\label{fig:acsedcompare}
\end{center}
\end{minipage}
\end{figure*}

\begin{figure*}
\begin{minipage}{7in}
\begin{center}
\includegraphics[width=0.99\linewidth]{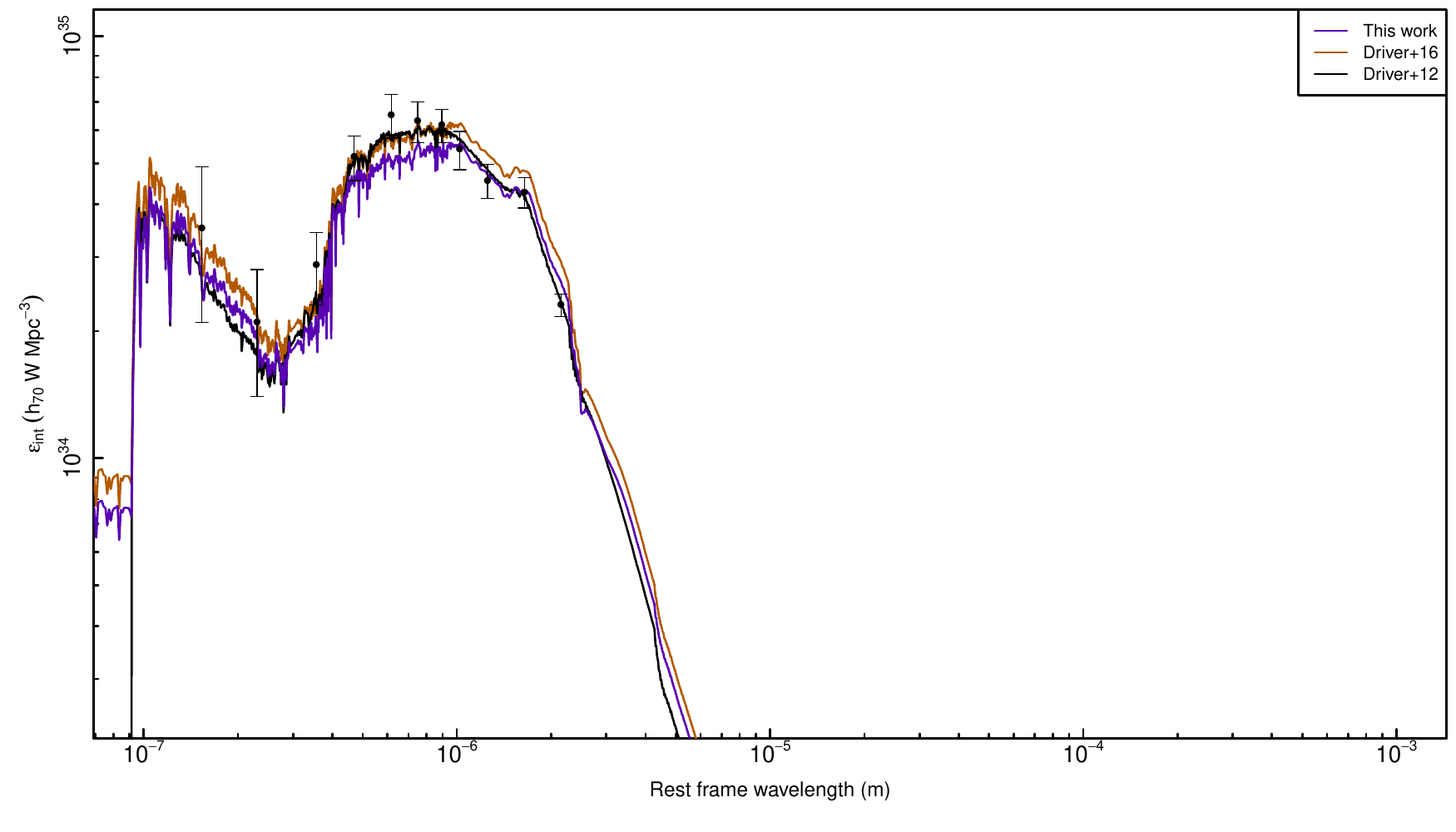}
\caption{The unattenuated CSEDs for the GAMA dataset compared to previous estimates for $0.02 < z < 0.08$. The black line and points represent the \citet{driver12} data and the orange line represents \citet{driver15}. The curves are subject to the errors described in Table \ref{tab:errors}.}
\label{fig:ucsedcompare}
\end{center}
\end{minipage}
\end{figure*}

\begin{table}
\begin{center}
\caption{Energy output as a function of redshift computed by integrating under the (attenuated) CSED over the full wavelength range and between 100 nm and $8~\mu$m (optical only). The quoted errors are derived from Table \ref{tab:errors}.}
\label{tab:csed}
\begin{tabular}{l|cc|}
\hline
Redshift      & Full CSED & Optical only \\
& \multicolumn{2}{c}{($10^{35}~h_{70}$~W~Mpc$^{-3}$)} \\
\hline
$0.02 - 0.08$ & $1.3 \pm 0.3$ & $0.8 \pm 0.2$ \\
$0.08 - 0.14$ & $1.6 \pm 0.2$ & $1.0 \pm 0.1$ \\
$0.14 - 0.20$ & $1.7 \pm 0.2$ & $1.0 \pm 0.1$ \\
$0.20 - 0.28$ & $1.9 \pm 0.4$ & $1.1 \pm 0.2$ \\
$0.28 - 0.36$ & $2.5 \pm 0.6$ & $1.4 \pm 0.4$ \\
$0.36 - 0.45$ & $2.5 \pm 0.7$ & $1.4 \pm 0.4$ \\
$0.45 - 0.56$ & $2.5 \pm 0.7$ & $1.2 \pm 0.3$ \\
$0.56 - 0.68$ & $3.3 \pm 0.8$ & $1.5 \pm 0.4$ \\
$0.68 - 0.82$ & $4.0 \pm 0.9$ & $1.8 \pm 0.4$ \\
$0.82 - 0.99$ & $5.1 \pm 1.0$ & $2.2 \pm 0.4$ \\
\hline
\end{tabular}
\end{center}
\end{table}

Integrating under our CSEDs gives the expected trend of declining energy output (Figure \ref{fig:summary}), consistent with star formation activity winding down since cosmic noon at $z = 1$ \citep{madau14}. This smooth decline can clearly be seen in the unattenuated ultraviolet emission in Figure \ref{fig:ucsed}. The unattenuated optical emission declines at a slower pace -- a result of the average age of stellar populations increasing, causing the unattenuated CSED to become redder.

For the GAMA sample, we measure a 24 per cent decline in energy output (see Figure \ref{fig:summary} and Table \ref{tab:csed}). This decline is significant at the 1$\sigma$ level to cosmic variance (CV) and other sources of error. Our low redshift results are noticeably lower than \citet{driver15}, as shown in Figures \ref{fig:acsedcompare} and \ref{fig:ucsedcompare}. This is for two reasons -- firstly, the more accurate aperture definitions and deblending of \citet{wright16a} have resulted in the exclusion of some stellar flux. Secondly, our correction factors for a given redshift bin depend only on that bin. This is more robust than the correction factor used by \citet{driver15} derived from dividing the total CSED and the CSED for all galaxies above a stellar mass threshold of $10^{10} M_\odot$.

At intermediate redshifts, the picture becomes slightly more complicated due to the uncertainty in the far-infrared portion of the CSED. We tentatively report a halving of energy output. This is robust to CV, but marginally significant with respect to uncertainty caused by the lack of measurements in the far-infrared. 

One exception is the decrease in energy output at optical wavelengths between the $0.14 < z < 0.20$ and $0.45 < z < 0.56$ bins. This we measure as the total energy output over the rest frame wavelength range $100$~nm~$< \lambda < 8~\mu$m with increasing redshift. The decline at intermediate redshifts is not significant with respect to CV. Ultraviolet and far-infrared emission, where significant extrapolation is present, continue to increase at this epoch, as shown in Figure \ref{fig:acsed}. 

Overall, the approximately factor four decline -- from $5.1 \pm 1.0 \times 10^{35}~h_{70}$~W~Mpc$^{-3}$ at $z = 0.91$ to $1.3 \pm 0.3 \times 10^{35}~h_{70}$~W~Mpc$^{-3}$ at $z = 0.06$ is robust to all forms of error. The decline in the total integrated energy output ($E = \int \frac{\epsilon}{\lambda} d\lambda$) is well fit by a smooth power law:

\begin{equation}
\log(E) = (2.18 \pm 0.12) \log(1+z) + (0.07 \pm 0.02) + 35
\end{equation}

This decline occurs at a slightly slower rate than the cosmic star formation density (see Figure \ref{fig:summary}). This is expected, as the energy density reflects not only current star formation, but also remaining stellar populations from previous star formation. The size of this effect is small -- resulting in a difference of a factor of $\sim$1.25 over the redshift range -- indicating that the energy output of the Universe is still dominated by young stars. Note that the cosmic star formation density has been arbitrarily scaled to match the above CSED fit in the highest redshift bin. Beyond $z = 1$ (not shown on Figures \ref{fig:acsed}, \ref{fig:ucsed} and \ref{fig:summary}), the CSED unphysically declines rapidly due to incompleteness in the \citet{andrews16} catalogue. A deep, near-infrared selected catalogue is required to probe to higher redshifts.

\subsection{The integrated photon escape fraction}

\begin{figure}
\begin{center}
\includegraphics[width=0.99\linewidth]{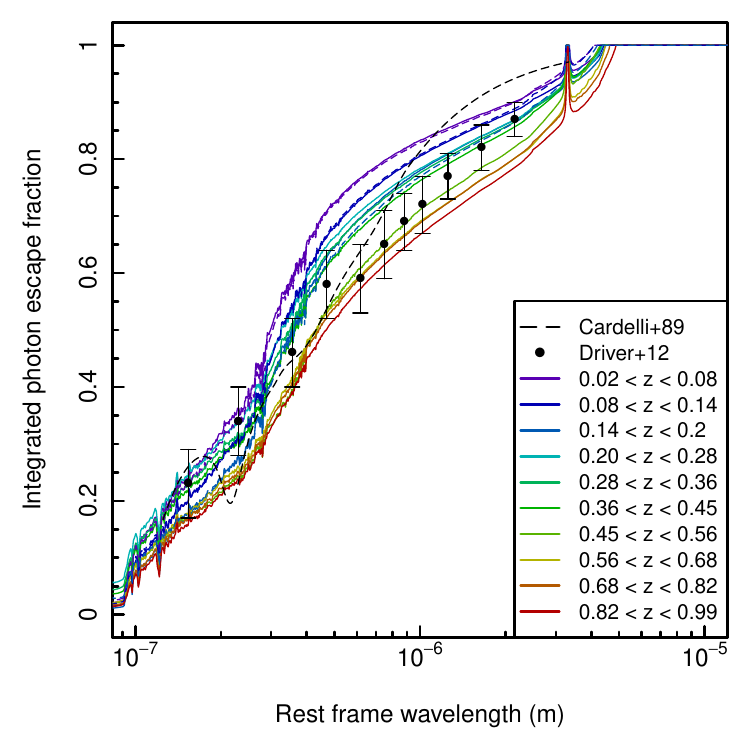}
\caption{The integrated photon escape fraction for the GAMA and G10 datasets. The IPEFs estimated by \citet{driver12} (grey line and black points) and \citet{driver15} (dashed) are also shown as is the average Milky Way attenuation curve \citep{cardelli89} assuming an escape fraction of 0.6 at 549.5~nm. Curves are subject to CSED shape errors as described in Section \ref{sec:csedbad}.}
\label{fig:ipef}
\end{center}
\end{figure}

\begin{figure}
\begin{center}
\includegraphics[width=0.99\linewidth]{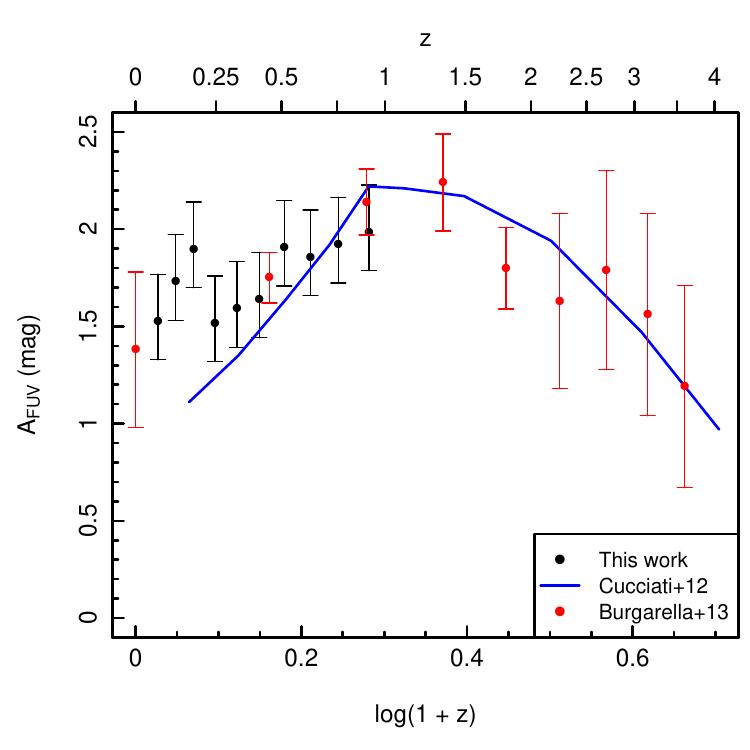}
\caption{$A_\mathrm{FUV}$ as a function of redshift (computed from convolving the IPEF curves with the \textsc{GALEX} FUV filter, with a 20 per cent indicative error in the IPEF) compared to \citet{cucciati12} and \citet{burgarella13}.}
\label{fig:extinction}
\end{center}
\end{figure}

Dividing the attenuated CSED by the unattenuated CSED gives the integrated photon escape fraction (IPEF), which we show in Figure \ref{fig:ipef}. Also shown in Figure \ref{fig:ipef} are the \citet{driver12} values adopted from the Millenium Galaxy Catalogue data spanning $0 < z < 0.18$ \citep{driver08}. The IPEF is a simple measure of the effect dust attenuation has on galaxy light. The IPEF is not subject to cosmic variance, jackknife, $L_\mathrm{opt}$  correction or photometric redshift errors as the uncertain normalization of the CSED at a given redshift is divided out. This leaves errors in the CSED shape only and dust attenuation as the dominant sources of uncertainty. These uncertainties are hard to estimate without a thorough evaluation of the SED modelling process, so we assign an indicative 20 per cent error to account for this. 

The IPEFs shown here are consistent with increasing opacity with lookback time, which presumably is linked to an increasing dust mass density. This is consistent with the findings of \citet{dunne11} and \citet{driver17} and not surprising given the correlation between star formation and dust opacity and the cosmic star formation density rising with redshift. This is also consistent with the evolution in the ultraviolet and infrared luminosity functions from $z > 1$ to $z = 0$ as noted by \citet{bernhard14}. The escape fraction at 150~nm decreases from $24 \pm 5$ per cent at $z = 0.05$ to $16 \pm 3$ per cent at $z = 0.915$. The escape fraction at 250~nm also declines from $40 \pm 8$ per cent to $26 \pm 5$ per cent over the same time intervals. This is in line with \citet{driver15}, with the caveat that the dust attenuation and re-emission curves may be unreliable as explained in Section \ref{sec:csedbad}. These attenuation curves are similar to that of the Milky Way \citep{cardelli89} when normalised to an escape fraction of 0.6 at 549.5~nm (black dashed line), with the lower redshift curves being more transparent. It is clear that further reduction in SED modelling errors is required to obtain a Universal extinction curve.

Convolving the IPEF with a filter curve and converting to magnitudes gives an attenuation coefficient. We show the resulting FUV attenuation coefficients ($A_\mathrm{FUV}$) with the 20 per cent indicative uncertainty in the IPEF in Figure \ref{fig:extinction}. Our estimates are in line with those derived from \citet{cucciati12} and \citet{burgarella13} using VVDS, PEP and HerMES data across the redshift range. 

\subsection{The integrated galactic light}

\begin{figure*}
\begin{minipage}{7in}
\begin{center}
\includegraphics[width=0.99\linewidth]{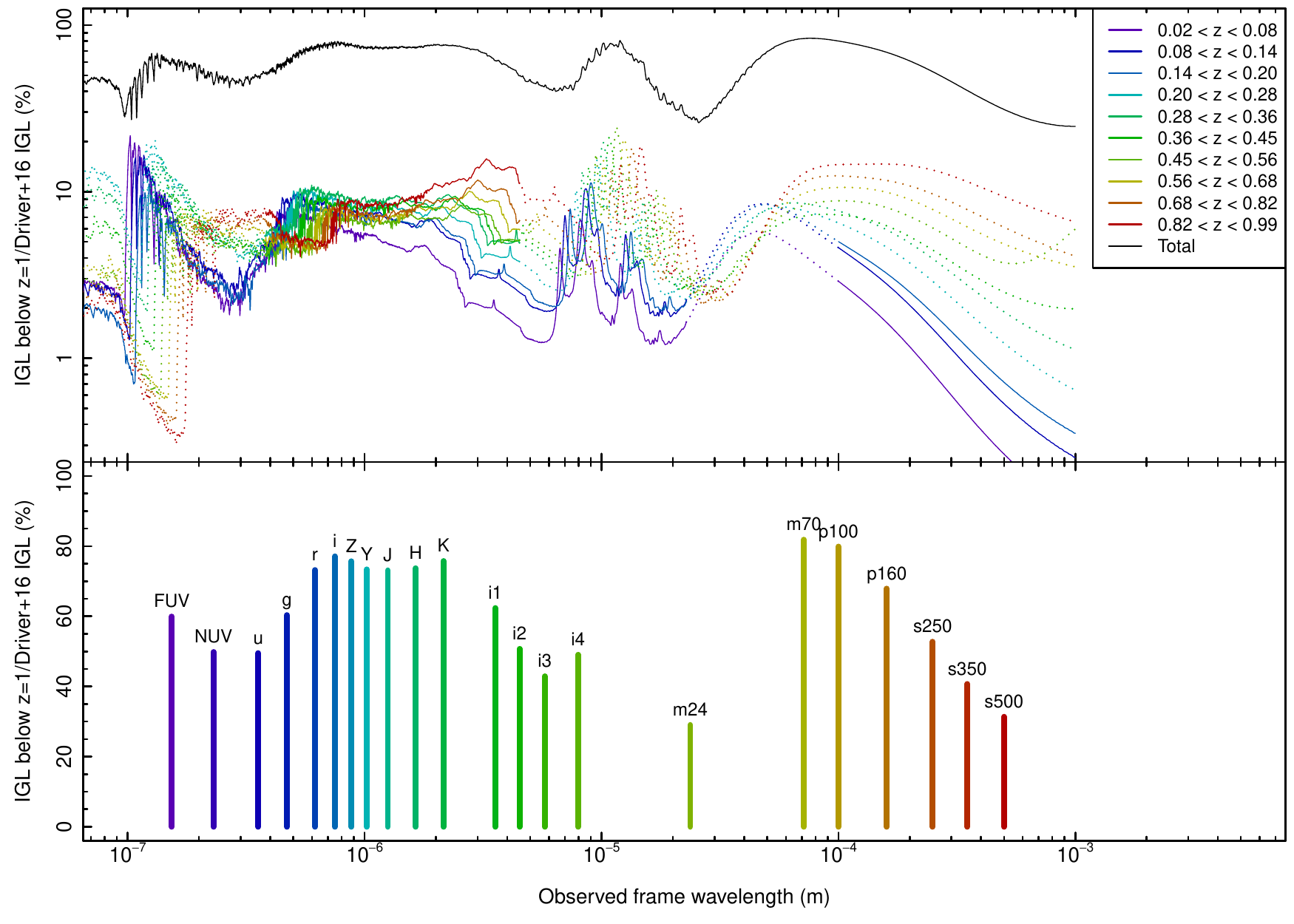}
\caption{The fraction of the total IGL, measured by the fitting function of \citet{driver16b} (top: as a function of wavelength and redshift, bottom: per photometric filter from \textit{GALEX}, SDSS, VISTA, IRAC, MIPS, PACS and SPIRE and summed over all redshifts), contributed by our CSEDs. Measurements are subject to the errors described in Table \ref{tab:errors} and the CSED shape errors described in Section \ref{sec:csedbad}.}
\label{fig:igl}
\end{center}
\end{minipage}
\end{figure*}

\begin{figure*}
\begin{minipage}{7in}
\begin{center}
\includegraphics[width=0.99\linewidth]{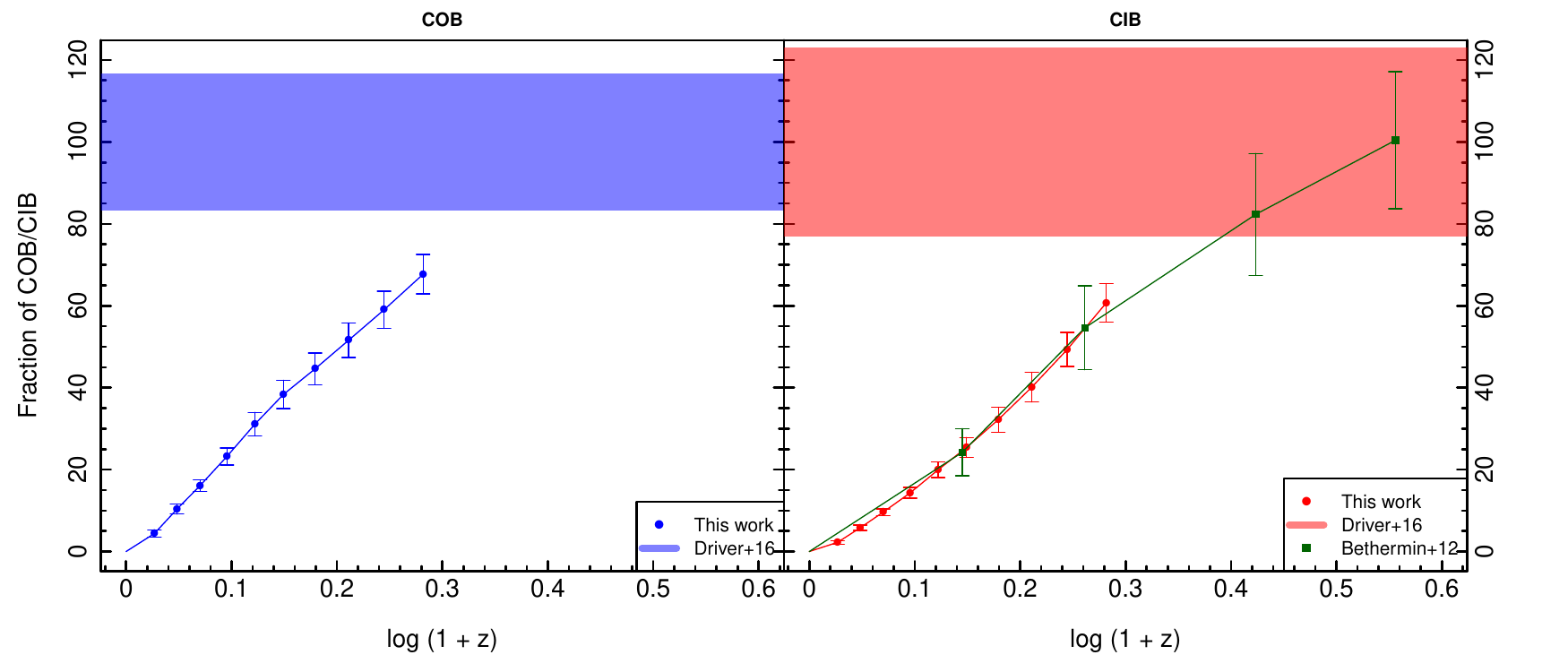}
\caption{The portion of the \citet{driver16b} cosmic optical (left) and infrared (right) backgrounds recovered as a function of redshift compared against \citet{bethermin12}. The shaded areas represent the uncertainty range of the \citet{driver16b} cosmic optical and infrared background measurements.}
\label{fig:cobfraction}
\end{center}
\end{minipage}
\end{figure*}

\begin{table}
\begin{center}
\caption{Contributions to the cosmic optical and infrared backgrounds (COB, CIB) and IGL as a function of redshift. Quoted uncertainties are derived from Table \ref{tab:errors}.}
\label{tab:igl}
\begin{tabular}{l|ccc|}
\hline
Redshift      & COB & CIB & IGL \\
& \multicolumn{3}{c}{(nW~m$^{-2}$~sr$^{-1}$)} \\
\hline
$0.02 - 0.08$ & $ 1.1 \pm 0.2$ & $ 0.6 \pm 0.1$ & $ 1.6 \pm 0.3$ \\
$0.08 - 0.14$ & $ 1.4 \pm 0.2$ & $ 0.9 \pm 0.1$ & $ 2.4 \pm 0.3$ \\
$0.14 - 0.20$ & $ 1.4 \pm 0.2$ & $ 1.0 \pm 0.1$ & $ 2.4 \pm 0.3$ \\
$0.20 - 0.28$ & $ 1.7 \pm 0.4$ & $ 1.2 \pm 0.3$ & $ 2.9 \pm 0.6$ \\
$0.28 - 0.36$ & $ 1.9 \pm 0.5$ & $ 1.5 \pm 0.4$ & $ 3.4 \pm 0.8$ \\
$0.36 - 0.45$ & $ 1.7 \pm 0.5$ & $ 1.4 \pm 0.4$ & $ 3.2 \pm 0.9$ \\
$0.45 - 0.56$ & $ 1.5 \pm 0.4$ & $ 1.8 \pm 0.5$ & $ 3.3 \pm 0.9$ \\
$0.56 - 0.68$ & $ 1.7 \pm 0.4$ & $ 2.1 \pm 0.5$ & $ 3.7 \pm 0.9$ \\
$0.68 - 0.82$ & $ 1.8 \pm 0.4$ & $ 2.4 \pm 0.5$ & $ 4.2 \pm 0.9$ \\
$0.82 - 0.99$ & $ 2.0 \pm 0.4$ & $ 3.0 \pm 0.6$ & $ 5.0 \pm 1.0$ \\
\hline
TOTAL         & $16.2 \pm 1.2$ & $15.8 \pm 1.2$ & $32.0 \pm 2.4$ \\
\hline
\end{tabular}
\end{center}
\end{table}

The integrated galactic light (IGL) at $z=0$ can be derived from the CSED $\epsilon$ as follows:

\begin{equation}
\mathrm{IGL}(\lambda) = \sum_{z=0}^{z=\infty} \frac{\epsilon(\lambda(1+z), z) dV(z)}{4 \pi d_l(z)^2 }
\end{equation}

where $d_l(z)$ is the luminosity distance and $dV(z)$ is the differential comoving volume of each redshift bin subtending a solid angle of 1~sr. Figure \ref{fig:igl} shows the fraction of the \citet{driver16b} IGL accounted for by redshifting and summing the CSEDs measured in this work. Our CSEDs constitute a roughly constant 40--80 per cent of the IGL across the entire wavelength range, with contributions to the cosmic optical background (the portion of the IGL between 100 nm and $8~\mu$m), cosmic infrared background (8~$\mu$m $< \lambda < 1~$mm) and IGL as a function of redshift shown in Table \ref{tab:igl}. Figure \ref{fig:cobfraction} shows the fraction of the cosmic optical and infrared backgrounds accounted for by emission prior to the given redshift.

As expected, low redshifts dominate FUV emission and high redshifts dominate in the mid-infrared and longwards of the cold dust peak. We do not recover the total IGL at any wavelength, indicating a significant portion was emitted before $z = 1$. Areas with relatively low level of IGL recovery should be easily explained by the high-redshift galaxy population. 

Contributions to the cosmic optical background (see Table \ref{tab:igl}) are approximately constant as a function of redshift, while contributions to the cosmic infrared background increase. For comparison, \citet{driver16b} measured cosmic optical and infrared backgrounds of $24 \pm 4$ and $26 \pm 5$~nW~m$^{-2}$~sr$^{-1}$ respectively, giving the shaded areas in Figure \ref{fig:cobfraction}. Our measurements are in excellent agreement with contributions to the cosmic infrared background measured by \citet{bethermin12} (green points/lines).

The combination of WAVES and a high-redshift sample based on \textit{James Webb Space Telescope} (\textit{JWST}), \textit{Wide Field Infrared Survey Telescope} (\textit{WFIRST}) and \textit{Euclid} data will allow the determination of the optical CSED out to the epoch of reionisation and thus the full characterisation of the IGL.

\section{Conclusion}
\label{sec:conclusion}

In this work, we measured the CSED for $0 < z < 1$ by stacking SED fits from \citet{driver17}. We binned the SEDs into 10 different redshift intervals and measured the CSED and integrated photon escape fraction for each through stacking. We found that energy output declines as a function of redshift, from $(5.1 \pm 1.0) \times 10^{35} h_{70}$~W~Mpc$^{-3}$ to $(1.3 \pm 0.3) \times 10^{35} h_{70}$~W~Mpc$^{-3}$ (Figure \ref{fig:summary}). This decline is robust despite cosmic variance (CV) and other uncertainties and occurs at a rate slightly slower than the decline in cosmic star formation. Combined with the reddening of the unattenuated CSED, this is consistent with the mean age of stellar populations becoming older. AGN do not contribute significantly to the CSED at any $z < 1$. We also show that the photon escape fraction also declines with increasing redshift, equivalent to an increase in $A_\mathrm{FUV}$ of 0.4~mag, consistent with an increase in the cosmic dust density. 

The CSEDs presented here complement the recent measurements of the integrated galactic light reported by \citet{driver16b}. We will follow up the CSEDs presented in this work with a comparison to semi-analytic (e.g. \citealt{gilmore12}) and empirical models (e.g. \citealt{driver13}) of the CSED and integrated galactic light as a function of cosmic time (Andrews et al. in prep). We will also explore the physical properties obtained in our SED fits for various (sub-)populations of galaxies (e.g. \citealt{kelvin14}) and investigate the biases inherent in the CSEDs in more detail. Finally, we will compare our physical parameters with measurements made using independent methods derived from the GAMA database and use them to improve the SED fitting process.

The dominant source of directly quantifiable error in our CSED estimates is CV. Incompleteness remains a significant impediment towards fully characterising the CSED over cosmic time. While the optical (100~nm $< \lambda < 8~\mu$m) portion of the CSED is well constrained, the relatively low sensitivity and resolution of the far-infrared data prevents the full characterization of the CSED at all redshifts. Incompleteness also causes a systematic underestimation of the ultraviolet and far-infrared flux of 10--30 per cent, as low luminosity galaxies are preferentially star forming and dusty. Systematic errors also arise from SED modelling and the non-modelling of AGN.

Data from next-generation surveys of galaxy evolution, such as WAVES and the Maunakea Spectroscopic Explorer, will significantly reduce incompleteness and CV. These surveys aim to cover hundreds of square degrees of sky to 2-4 magnitudes fainter than GAMA with a high level of spectroscopic completeness. In addition, this will enable us to replace photometric redshifts with spectroscopic data for a significant portion of the sample. The large survey footprint will also reduce the uncertainty in the CSED normalization due to cosmic variance -- for example the 26 per cent CV in the 0.915 deg$^2$ observable portion of the G10 field for $0.45 < z < 0.56$ will be reduced to 7 per cent for an illustrative survey of two 25 deg$^2$ fields. The combination of WAVES with observations equivalent to COSMOS using \textit{JWST} and \textit{WFIRST} will, once combined using the procedure in this work, constrain the ultraviolet to near-infrared CSED out to the epoch of reionisation and thus how the cosmic optical background builds up over the history of the Universe.

\section*{Acknowledgements}
We thank the referee, Matthieu B\'{e}thermin, for a prompt report that improved this paper. SKA is supported by the Australian Government's Department of Industry Australian Postgraduate Award (APA). 

GAMA is a joint European-Australasian project based around a spectroscopic campaign using the Anglo-Australian Telescope. The GAMA input catalogue is based on data taken from the SDSS and the UKIRT Infrared Deep Sky Survey. Complementary imaging of the GAMA regions is being obtained by a number of independent survey programmes including GALEX MIS, VST KiDS, VISTA VIKING, WISE, Herschel-ATLAS, GMRT and ASKAP providing UV to radio coverage. GAMA is funded by the STFC (UK), the ARC (Australia), the AAO, and the participating institutions. The GAMA website is \url{http://www.gama-survey.org/}. Based on observations made with ESO Telescopes at the La Silla Paranal Observatory under programme ID 179.A-2004.

The G10/COSMOS redshift catalogue and cutout tool uses data acquired as part of the Cosmic Evolution Survey (COSMOS) project and spectra from observations made with ESO Telescopes at the La Silla or Paranal Observatories under programme ID 175.A-0839. The G10 cutout tool is hosted and maintained by funding from the International Centre for Radio Astronomy Research (ICRAR) at the University of Western Australia.

This work is supported by resources provided by the Pawsey Supercomputing Centre with funding from the Australian Government and the Government of Western Australia.

\label{lastpage}

\section*{Supporting information}
The attenuated and unattenuated rest frame CSEDs in $\lambda f_\lambda$ units are available online as supporting information. The first five rows of the lowest redshift CSED, \textit{z002-008\_corr.csed\_norm}, are given in Table \ref{tab:supplementary}.

\begin{table}
\begin{center}
\caption{Extract from the lowest redshift CSED. The full tables for all redshifts are available as supporting information.}
\label{tab:supplementary}
\begin{tabular}{|ccc|}
\hline
Wavelength & Attenuated  & Unattenuated  \\
(m)        & CSED (W)    & CSED (W) \\
\hline
1e-08      & 2.7749e+28 & 5.1875e+31 \\
1.0009e-08 & 2.8026e+28 & 5.2347e+31 \\
1.0018e-08 & 2.8269e+28 & 5.2533e+31 \\
1.0027e-08 & 2.8512e+28 & 5.2707e+31 \\
1.0036e-08 & 2.8755e+28 & 5.2883e+31 \\
\hline
\end{tabular}
\end{center}
\end{table}

\begin{thebibliography}{99}
\bibitem[\protect\citeauthoryear{Abazajian et al.}{2009}]{abazajian09} Abazajian K.~N. et al., 2009, ApJS, 182, 543
\bibitem[\protect\citeauthoryear{Ahn et al.}{2014}]{ahn14} Ahn C.~P. et al., 2014, ApJS, 211, 17
\bibitem[\protect\citeauthoryear{Ahnen et al.}{2016}]{magic16} Ahnen M.~L. et al. 2016, A\&A, 590, A24
\bibitem[\protect\citeauthoryear{Andrews et al.}{2017}]{andrews16} Andrews S.~K., Driver S.~P., Davies L.~J~.M., Kafle P.~R., Robotham A.~S.~G., Wright A.~H. 2017, MNRAS, 464, 1569
\bibitem[\protect\citeauthoryear{Bernhard et al.}{2014}]{bernhard14} Bernhard E., B\'{e}thermin M., Sargent M., Buat V., Mullaney J.~R., Pannella M., Heinis S., Daddi E. 2014, MNRAS, 442, 509
\bibitem[\protect\citeauthoryear{Berta et al.}{2011}]{berta11} Berta S. et al. 2011, A\&A 532, A49
\bibitem[\protect\citeauthoryear{B\'{e}thermin et al.}{2010}]{bethermin10} B\'{e}thermin M., Dole H., Beelan A., Aussel H. 2010, A\&A, 512, A78
\bibitem[\protect\citeauthoryear{B\'{e}thermin et al.}{2012a}]{bethermin12} B\'{e}thermin M., et al. 2012a, A\&A, 542, A58
\bibitem[\protect\citeauthoryear{B\'{e}thermin et al.}{2012b}]{bethermin12b} B\'{e}thermin M. et al. 2012b, ApJ, 757, L23
\bibitem[\protect\citeauthoryear{Burgarella et al.}{2013}]{burgarella13} Burgarella D. et al. 2013, A\&A, 554, A70
\bibitem[\protect\citeauthoryear{Bruzual}{2007}]{bruzual07} Bruzual, G. 2007, in ASP Conf. Ser. 374, From Stars to Galaxies: Building the Pieces to Build Up the Universe, ed. A. Vallenari et al. (San Francisco, CA: ASP), 303
\bibitem[\protect\citeauthoryear{Bruzual \& Charlot}{2003}]{bc03} Bruzual G. \& Charlot S. 2003, MNRAS, 344, 1000
\bibitem[\protect\citeauthoryear{Bruzual et al.}{2013}]{bruzual13} Bruzual, G., Charlot, S., Gonz\'{a}lez L\'{o}pezlira, R., Srinivasan S., Boyer M.~L., Riebel D. 2013, in Proceedings of the IAU Symposium No. 295 ``The LF of TP-AGB stars in the LMC/SMC" eds. D. Thomas, A. Pasquali \& I. Ferreras, Cambridge: Cambridge University Press, 282
\bibitem[\protect\citeauthoryear{Capak et al.}{2007}]{capak07}  Capak P. et al. 2007, ApJS, 172, 284
\bibitem[\protect\citeauthoryear{Capozzi et al.}{2016}]{capozzi16} Capozzi D., Maraston C., Daddi E., Renzini A., Strazzullo V., Gobat R. 2016, MNRAS, 456, 790
\bibitem[\protect\citeauthoryear{Cardelli et al.}{1989}]{cardelli89} Cardelli J.~A., Clayton G.~C., Mathis J.~S. 1989, ApJ, 345, 245
\bibitem[\protect\citeauthoryear{Carniani et al.}{2015}]{carniani15} Carniani S. et al. 2015, A\&A, 584, A78
\bibitem[\protect\citeauthoryear{Chabrier}{2003}]{chabrier03} Chabrier G. 2003, PASP, 115, 763
\bibitem[\protect\citeauthoryear{Charlot \& Fall}{2000}]{charlot00} Charlot S. \& Fall S.~M. 2000, ApJ, 539, 718
\bibitem[\protect\citeauthoryear{Coil et al.}{2011}]{coil11} Coil A.~L. et al., 2011, ApJ, 741, 8
\bibitem[\protect\citeauthoryear{Cool et al.}{2013}]{cool13} Cool R.~J. et al., 2013, ApJ, 767, 118
\bibitem[\protect\citeauthoryear{Conroy \& Gunn}{2010}]{conroy10} Conroy C. \& Gunn J.~E. 2010, ApJ, 712, 833
\bibitem[\protect\citeauthoryear{Cooray et al.}{2012a}]{cooray12b} Cooray A., Gong Y. Smidt J., Santos M.~G. 2012, ApJ, 756, 92
\bibitem[\protect\citeauthoryear{Cooray et al.}{2012b}]{cooray12} Cooray A. et al. 2012, Nature, 490, 514
\bibitem[\protect\citeauthoryear{Cucciati et al.}{2012}]{cucciati12} Cucciati O. et al. 2012, A\&A, 539, A31
\bibitem[\protect\citeauthoryear{da Cunha, et al.}{2008}]{dacunha08} da Cunha E., Charlot S., Elbaz D. 2008, MNRAS 388, 1595
\bibitem[\protect\citeauthoryear{Dale \& Helou}{2002}]{dale02} Dale D. \& Helou G., 2002, ApJ, 576, 159
\bibitem[\protect\citeauthoryear{Darvish et al.}{2017}]{darvish17} Darvish B., Sobral D., Mobasher B., Scoville N.~Z., Best P., Sales L.~V., Smail I. 2017, ApJ, 796, 51
\bibitem[\protect\citeauthoryear{Davies et al.}{2015}]{davies15} Davies L.~J.~M. et al. 2015, MNRAS, 447, 1014
\bibitem[\protect\citeauthoryear{Davies et al.}{2016}]{davies16} Davies L.~J.~M. et al. 2016, MNRAS, 461, 458
\bibitem[\protect\citeauthoryear{Dom\'{i}nguez et al.}{2011}]{dominguez11} Dom\'{i}nguez A. et al. 2011, MNRAS, 410, 2556
\bibitem[\protect\citeauthoryear{Donley et al.}{2012}]{donley12} Donley J.~L. et al. 2012, ApJ, 748, 142
\bibitem[\protect\citeauthoryear{Driver \& Robotham}{2010}]{driver10} Driver S.~P. \& Robotham A.~S.~G. 2010, MNRAS, 407, 2131
\bibitem[\protect\citeauthoryear{Driver et al.}{2008}]{driver08} Driver S.~P., Popescu C.~C., Tuffs R.~J., Graham A.~W., Liske J., Baldry I.~K. 2008, ApJ, 678, L101
\bibitem[\protect\citeauthoryear{Driver et al.}{2011}]{driver11} Driver S.~P. et al. 2011, MNRAS, 413, 971
\bibitem[\protect\citeauthoryear{Driver et al.}{2012}]{driver12} Driver S.~P. et al. 2012, MNRAS, 427, 3244
\bibitem[\protect\citeauthoryear{Driver et al.}{2013}]{driver13} Driver S.~P., Robotham A.~S.~G., Bland-Hawthorn J., Brown M., Hopkins A., Liske J., Phillipps S., Wilkins S. 2013, MNRAS, 430, 2622
\bibitem[\protect\citeauthoryear{Driver et al.}{2016a}]{waves} Driver S.~P., Davies L.~J., Meyer M., Power C., Robotham A.~S.~G., Baldry I.~K., Liske J., Norberg P. 2016, ASSP, 42, 205
\bibitem[\protect\citeauthoryear{Driver et al.}{2016b}]{driver15} Driver S.~P. et al. 2016b, MNRAS, 455, 3911
\bibitem[\protect\citeauthoryear{Driver et al.}{2016c}]{driver16b} Driver S.~P. et al. 2016c, ApJ, 827, 108
\bibitem[\protect\citeauthoryear{Driver et al.}{2017}]{driver17} Driver S.~P. et al. 2017, MNRAS, submitted
\bibitem[\protect\citeauthoryear{Dunne et al.}{2011}]{dunne11} Dunne L. et al. 2011, MNRAS, 417, 1510
\bibitem[\protect\citeauthoryear{Dwek et al.}{1998}]{dwek98} Dwek E., et al. 1998, ApJ, 508, 106
\bibitem[\protect\citeauthoryear{Eales et al.}{2010}]{eales10} Eales S. et al., 2010, PASP, 122, 499
\bibitem[\protect\citeauthoryear{Edge et al.}{2013}]{edge13} Edge A., Sutherland W., Kuijken K., Driver S.~P., McMahon R., Eales S., Emerson J.~P. 2013, The Messenger, 154, 32
\bibitem[\protect\citeauthoryear{Fazio et al.}{2004}]{fazio04} Fazio G.~G. et al. 2004, ApJS, 154, 39
\bibitem[\protect\citeauthoryear{Finke et al.}{2010}]{finke10} Finke J.~D., Razzaque S., Dermer C.~D. 2010, ApJ, 712, 238
\bibitem[\protect\citeauthoryear{Franceschini et al.}{2008}]{franceschini08} Franceschini A., Rodighiero G., Vaccari M. 2008, A\&A, 487, 837
\bibitem[\protect\citeauthoryear{Gardner et al.}{2000}]{gardner00} Gardner J.~P., Brown T.~M. \& Ferguson H.~C. 2000, ApJ, 542, L79
\bibitem[\protect\citeauthoryear{Garilli et al.}{2008}]{garilli08} Garilli B. et al., 2008, A\&A, 486, 683
\bibitem[\protect\citeauthoryear{Gil de Paz et al.}{2005}]{gildepaz05} Gil de Paz A. et al. 2005, ApJ, 627, L29
\bibitem[\protect\citeauthoryear{Gilmore et al.}{2012}]{gilmore12} Gilmore R.~C., Somerville R.~S., Primack J.~R., Dom\'{i}nguez A. 2012, MNRAS, 422, 3189
\bibitem[\protect\citeauthoryear{Hauser \& Dwek}{2001}]{hauser01} Hauser M.~G. \& Dwek E. 2001, ARA\&A, 39, 249
\bibitem[\protect\citeauthoryear{H.E.S.S. Collaboration}{2013}]{hess13} H.E.S.S. Collaboration 2015 A\&A, 550, A4
\bibitem[\protect\citeauthoryear{Ilbert et al.}{2009}]{ilbert09} Ilbert O. et al. 2009, ApJ, 690, 1236
\bibitem[\protect\citeauthoryear{Inoue et al.}{2013}]{inoue13} Inoue Y., Inoue S., Kobayashi M.~A.~R., Makiya R., Niino Y., Totani T. 2013, ApJ, 768, 197
\bibitem[\protect\citeauthoryear{Jauzac et al.}{2011}]{jauzac11} Jauzac M. et al. 2011, A\&A, 525, A52
\bibitem[\protect\citeauthoryear{Kelvin et al.}{2014}]{kelvin14} Kelvin L.~S. et al. 2014, MNRAS, 439, 1245
\bibitem[\protect\citeauthoryear{Kennicutt}{1998}]{kennicutt98} Kennicutt R.~C. 1998, ARA\&A, 36, 189
\bibitem[\protect\citeauthoryear{Laigle et al.}{2016}]{laigle16} Laigle C. et al. 2016, ApJS, 224, 24
\bibitem[\protect\citeauthoryear{Lilly et al.}{2007}]{lilly07} Lilly S.~J. et al. 2007, ApJS, 172, 70
\bibitem[\protect\citeauthoryear{Lilly et al.}{2009}]{lilly09} Lilly S.~J. et al. 2009, ApJS, 184, 218
\bibitem[\protect\citeauthoryear{Liske et al.}{2015}]{liske15} Liske J. et al. 2015, MNRAS, 452, 2087
\bibitem[\protect\citeauthoryear{Lutz et al.}{2011}]{lutz11} Lutz D. et al. 2011, A\&A, 532, A90
\bibitem[\protect\citeauthoryear{Madau \& Pozzetti}{2000}]{madau00} Madau P. \& Pozzetti L. 2000, MNRAS, 312, L9
\bibitem[\protect\citeauthoryear{Madau \& Dickinson}{2014}]{madau14} Madau P. \& Dickinson M. 2014, ARA\&A, 52, 415
\bibitem[\protect\citeauthoryear{Maraston}{2005}]{maraston05} Maraston C. 2005, MNRAS, 362, 799
\bibitem[\protect\citeauthoryear{Maraston et al.}{2006}]{maraston06} Maraston C., Daddi E., Renzini A., Cimatti A., Dickinson M., Papovich C., Pasquali A., Pirzkal N. 2006, ApJ, 652, 82
\bibitem[\protect\citeauthoryear{Marsden et al.}{2009}]{marsden09} Marsden G. et al. 2009, ApJ, 707, 1729
\bibitem[\protect\citeauthoryear{Martin et al.}{2005}]{martin05} Martin C. et al., 2005, ApJ, 619, 1
\bibitem[\protect\citeauthoryear{McCracken et al.}{2012}]{mccracken12} McCracken H.~J. et al. 2012, A\&A, 544, A156
\bibitem[\protect\citeauthoryear{No\"{e}l et al.}{2013}]{noel13} No\"{e}l N.~E.~D., Greggio L., Renzini A., Carollo C.~M., Maraston C. 2013, ApJ, 772, 58
\bibitem[\protect\citeauthoryear{Oliver et al.}{2012}]{oliver12} Oliver S.~J. et al. 2012, MNRAS, 424, 1614
\bibitem[\protect\citeauthoryear{Partridge \& Peebles}{1967a}]{partridge67a} Partridge R.~B. \& Peebles P.~J.~E. 1967a, ApJ, 147, 868
\bibitem[\protect\citeauthoryear{Partridge \& Peebles}{1967b}]{partridge67b} Partridge R.~B. \& Peebles P.~J.~E. 1967b, ApJ, 148, 377
\bibitem[\protect\citeauthoryear{Richards et al.}{2006}]{richards06} Richards G. et al., 2006, AJ, 131, 2766
\bibitem[\protect\citeauthoryear{Sanders et al.}{2007}]{sanders07} Sanders D.~B. et al. 2007, ApJS, 172, 86
\bibitem[\protect\citeauthoryear{Scoville et al.}{2007}]{scoville07} Scoville N. et al. 2007, ApJS, 172, 1
\bibitem[\protect\citeauthoryear{Somerville et al.}{2012}]{somerville12} Somerville R.~S., Gilmore R.~C., Primack J.~R., Dom\'{i}nguez A. 2012, MNRAS, 423, 1992
\bibitem[\protect\citeauthoryear{Symeonidis et al.}{2013}]{symeonidis13} Symeonidis M. et al. 2013, MNRAS, 431, 2317
\bibitem[\protect\citeauthoryear{Taniguchi et al.}{2007}]{taniguchi07} Taniguchi Y. et al. 2007, ApJS, 192, 9
\bibitem[\protect\citeauthoryear{Taniguchi et al.}{2015}]{taniguchi16} Taniguchi Y. et al. 2015, PASJ, 67, 104
\bibitem[\protect\citeauthoryear{Taylor et al.}{2011}]{taylor11} Taylor E.~N. et al. 2011, MNRAS, 418, 1587
\bibitem[\protect\citeauthoryear{Thilker et al.}{2007}]{thilker07} Thilker D.~A. et al. 2007, ApJS, 173, 538
\bibitem[\protect\citeauthoryear{Viero et al.}{2013}]{viero13} Viero M.~P. et al. 2013, ApJ, 779, 32
\bibitem[\protect\citeauthoryear{Wright et al.}{2010}]{wright10} Wright E.~L. et al., 2010, AJ, 140, 1868
\bibitem[\protect\citeauthoryear{Wright et al.}{2016a}]{wright16a} Wright A.~H. et al. 2016, MNRAS, 460, 765
\bibitem[\protect\citeauthoryear{Xu et al.}{2005}]{xu05} Xu C. et al. 2005, ApJ, 619, L11
\bibitem[\protect\citeauthoryear{Zamojski et al.}{2007}]{zamojski07} Zamojski M.~A. et al. 2007, ApJ, 172, 468
\bibitem[\protect\citeauthoryear{Zemcov et al.}{2014}]{zemcov14} Zemcov M. et al. 2014, Science, 346, 732

\end{thebibliography}
\end{document}